%% file: durable-goods.tex
\documentclass[preprint,12pt,authoryear,review]{elsarticle} 

\usepackage{amssymb,amsmath,amsthm} 
\usepackage{mathtools} 
\usepackage{algorithm,algcompatible} 
\usepackage{comment} 
\usepackage{setspace} 
\usepackage{booktabs} 
\usepackage{bm} 
\usepackage[dvipsnames]{xcolor} 
\usepackage[left=1in, right=1in]{geometry} 
\usepackage[utf8]{inputenc} 
\usepackage[T2A,T1]{fontenc} 
\usepackage[english]{babel} 
\usepackage{array} 
\usepackage{multirow} 
\usepackage{caption} 
\usepackage{threeparttable} 
\usepackage[normalem]{ulem} 

\input{newcommands.tex}

\begin{document} 

\begin{frontmatter} 

\title{Optimal Investment, Consumption, and Insurance \\ with Durable Goods under Stochastic Depreciation Risk} 



\author[zurichethaddr,viennawuaddr]{Aleksandar~Arandjelovi\'{c}} 
\author[mqaddr]{Ryle~S.~Perera} 
\author[mqaddr1]{Pavel~V.~Shevchenko\corref{corrauth}} 
\author[mqaddr1]{Tak~Kuen~Siu} 
\author[utsaddr2,mqaddr1]{Jin~Sun} 

\cortext[corrauth]{Corresponding author: pavel.shevchenko@mq.edu.au} 

\address[zurichethaddr]{Department of Mathematics, ETH Zurich, Switzerland} 
\address[viennawuaddr]{Institute for Statistics and Mathematics, Vienna University of Economics and Business, Austria} 
\address[mqaddr]{Department of Applied Finance, Macquarie University, Australia} 
\address[mqaddr1]{Department of Actuarial Studies and Business Analytics, Macquarie University, Australia} 
\address[utsaddr2]{Faculty of Sciences, University of Technology Sydney, Australia} 

\singlespacing 

\begin{abstract} 
We study an infinite-horizon optimal investment, consumption and insurance problem for an economic agent who consumes a perishable and a durable good. 
The agent trades in a risk-free asset, a risky asset, and a durable good whose price follows a correlated diffusion, while the stock of the durable good depreciates deterministically and is subject to insurable Poisson loss shocks. 
The agent can partially hedge these shocks via an insurance contract with loading and chooses optimal perishable consumption, portfolio holdings, and insurance coverage to maximise expected discounted CRRA utility. 
Exploiting the homogeneity of the problem, we reduce the Hamilton--Jacobi--Bellman equation to a static one-dimensional optimisation over constant portfolio shares and derive a semi-explicit optimal strategy. 
We then prove a verification theorem for the associated jump-diffusion wealth process with insurance, establishing the existence and optimality of this constant-fraction strategy under explicit transversality conditions for both risk-aversion regimes $0<\gamma<1$ and $\gamma>1$. 
Numerical experiments illustrate the impact of stochastic depreciation risk and insurance loading on the optimal allocation to financial assets, durable goods, and insurance coverage. 
\end{abstract} 



\end{frontmatter} 

\clearpage 

\section{Introduction}
Durable consumption goods, such as houses or automobiles, are exposed to damages from events like natural disasters. 
A natural response for a risk-averse agent is to manage such losses by purchasing insurance, thereby exchanging the random loss for a certain premium. 
In this way, the potential loss may be minimized by simply paying a premium in advance to the insurer who will compensate the agent in the event of a loss or accident (i.e.\ negative monetary shocks). 
Although such arrangements are common, the insurance premiums are costly, and they reduce the utility that can be derived from saving or consuming. 
This motivates the agent to follow a self-insurance strategy by accumulating buffer wealth. 
However, absorbing a large loss in the absence of insurance could decrease utility, current consumption, and future utility from bequests. 
\citet{Mossin1968} showed that full insurance coverage is not optimal when the premium includes a positive loading. 
\citet{arrow1971} showed that it is optimal to purchase a deductible insurance contract on aggregate wealth when a positive loading premium is charged. 
This strand of classic literature is concerned with the case of a single insurable asset in a static model setting without considering consumption. 
Therefore, the agent implicitly transforms the corresponding loss into reduced consumption or savings. 
Continuous-time models capture the interaction between the risk of loss and consumption. 
In a dynamic setting extending the analysis in \citet{Briys1986}, \citet{moore2006} showed that an optimal insurance is a deductible insurance for a risk-averse investor. 
\citet{Perera2010} examined the demand for insurance in a continuous-time modelling framework by applying the martingale approach to \citet{moore2006}. 


Modern economies offer a wide variety of consumption goods. 
For the purpose of modelling, each good can be classified as either a \emph{perishable} good or a \emph{durable} good. 
Insurable assets are often durable consumption goods, such as houses or automobiles. 
A durable good such as a house differs from other financial assets (e.g.\ stocks and bonds), as the owner can derive both utility and home equity from it. 
For many economic agents, a house is the largest asset in their portfolio. 
However, the interaction of such a positive durable asset with the other financial assets has not yet received sufficient attention in the literature on the optimal investment-consumption-insurance choice. 
The main challenge is how to incorporate various frictions and factors. 
Trading durable consumption goods induces time-varying investment opportunities which affect investment strategies in financial assets. 


Numerous researchers have studied optimal consumption and investment with durable consumption goods. 
Some literature includes \citet{hindy1993}, \citet{detemple1996}, \citet{cuoco2000}, \citet{cocco2004}, \citet{damgaard2003} and \citet{matos2014}. 
However, these studies are limited by not incorporating insurance on durable consumption goods. 
This study incorporates the role of insurance on durable goods into a continuous-time framework for optimal consumption and investment. 
Specifically, in this paper, we develop a model to find not only optimal consumption and investment with perishable and durable goods but also optimal level of insurance on durable goods that can be damaged by insurable events. 




We formulate the agent’s optimisation problem as a stochastic control problem for a jump-diffusion wealth process. 
Then the effect of insurance costs is investigated by studying the optimal value function. 
Using the dynamic programming approach in stochastic optimal control, we derive the Hamilton--Jacobi--Bellman (HJB) equation which leads to a semi-explicit solution for the optimal insurance coverage. 
We also establish a verification theorem for the candidate value function and the candidate optimal strategy. 
Specifically, sufficient conditions are identified for the candidate value function and the candidate optimal strategy to be the value function and an optimal strategy, respectively. 
To this end, we adopt the approach in \cite{oksendal2019applied}. 
In \cite{damgaard2003}, a verification theorem for controlled diffusions was mentioned by referring to Theorem IV.5.1 in \cite{fleming1993controlled}. 
However, they did not discuss in detail the verification theorem for the candidate value function and optimal policies without transaction costs even in the case of controlled diffusions. 
While \cite{matos2014} incorporated jumps in the return process of a risky asset, they did not discuss a verification theorem for the candidate value function and optimal policies in the absence of transaction costs.
The verification theorem established in the current paper is general enough to incorporate the respective verification theorems for the models in \cite{damgaard2003} and \cite{matos2014} as special cases when transaction costs are not considered. 
Moreover, concavity of the HJB function was not proved for special cases considered in \cite{damgaard2003} and \cite{matos2014} but we prove it for our more general model. 

It appears that all the aforementioned literature only focused on the case $0<\gamma<1$, where $\gamma$ is the risk aversion parameter in the CRRA utility function $u(x)=x^{1-\gamma}/(1-\gamma)$. 
The empirically relevant case $\gamma > 1$ has been addressed only in a few papers. 
For example, \cite{davis1990portfolio} and \cite{herdegen2021elementary} study $\gamma > 1$ via perturbations of the candidate value function and of the utility function, respectively, while \cite{biffis2025high} explore the use of homogeneity arguments without explicit transversality conditions.
These contributions, however, are all set in diffusion models without jumps. 
In contrast, we consider the setting where jumps are present and provide a verification theorem for the case where the risk aversion parameter is greater than one. 
Specifically, the verification theorem for this case is based on the verification theorem in Theorem 5.1 of \cite{oksendal2019applied}. 
One of the key conditions to be verified is the uniform integrability of the candidate value function. 
This provides a direct and comparatively elementary approach to proving a verification theorem for the case where the risk aversion parameter exceeds one. 

We show that higher premium loadings reduce both the demand for durable goods and the associated insurance coverage. 
For sufficiently high premium loadings, it is optimal not to purchase insurance at all, which is intuitive. 
This result is consistent with \cite{arandjelovic2023life}, which shows that high loadings in life insurance and life annuities can generate periods of non-participation in a life-cycle insurance model. 
We also analyse the optimal solution with respect to changes in other model parameters, and we document how the optimal allocations react to variations in, for example, the jump intensity of insurable events, the loss fraction, and the risk aversion parameter. 

To summarise, our main contributions are as follows. 
\begin{enumerate} 
\item We incorporate insurable jump risk on the durable consumption good into the continuous-time consumption-investment problem with both perishable and durable goods, and we derive a semi-explicit constant-fraction solution in the frictionless case. 
\item We prove a verification theorem for the resulting jump-diffusion control problem with insurance, which yields sufficient conditions for optimality and specialises to the models of \cite{damgaard2003} and \cite{matos2014} when the additional insurance feature is switched off. 
\item In contrast to existing durable-good models that restrict attention to $0<\gamma<1$, we treat both $0<\gamma<1$ and the empirically relevant case $\gamma>1$, providing explicit transversality conditions that ensure optimality of the candidate strategy. 
\end{enumerate} 

The paper is structured as follows: 
First, the economic model is formally defined in Section~\ref{EconomicModel_sec}. 
Section~\ref{notransactioncos_sec} presents a semi-explicit solution. 
Section~\ref{numerical_sec} presents a numerical analysis of the solution. 
Finally, Section~\ref{conclusion_sec} summarizes the key findings and provides concluding remarks. 
The proofs of the results presented in the main text are provided in the appendices. 

\section{The model}\label{EconomicModel_sec}

We consider a continuous-time economy with an infinite-time horizon under which economic activities take place. 
To focus on the key elements and simplify the discussion, we suppose that an economic agent has access to investment and consumption opportunities including a perishable consumption good, a durable consumption good, and two financial securities subject to a positive discount rate in their valuations. 
In our model setup, there is no finite terminal date for investment decisions, and the economic agent is concerned with the dynamic behaviour of her portfolio over an infinite horizon. 

The resolution of uncertainties in the economy is described by a stochastic basis $(\Omega, \mathcal{F}, \mathbb{F}, \mathbb{P})$ such that $\mathbb{F} = (\mathcal{F}_{t})_{t \ge 0}$ is the $\mathbb{P}$-augmentation of the natural filtration generated by a two-dimensional Brownian motion $\{ (B_{1}(t), B_{2}(t)) \}_{t \ge 0}$ and a two-dimensional Poisson process $\{ (N_{1}(t), N_{2}(t)) \}_{t \ge 0}$ with constant intensities $\lambda_{1}$ and $\lambda_{2}$, respectively. 
Here, the processes $\{ B_{1}(t) \}_{t \ge 0}$, $\{ B_{2}(t) \}_{t \ge 0}$, $\{ N_{1}(t) \}_{t \ge 0}$ and $\{ N_{2} (t) \}_{t \ge 0}$ are assumed to be independent under the probability measure $\mathbb{P}$. 
Some assumptions are imposed in our stochastic dynamic modelling framework which are specified as follows: 
(a) There are no transaction costs or taxes incurred when trading assets. 
(b) Financial securities and durable consumption goods can be bought in unlimited quantities and in any (fractional) amounts. 
(c) Financial securities can be shorted, while durable consumption goods cannot be sold short. 

We assume, as in \cite{matos2014}, that the financial market consists of a non-dividend-paying risky asset $S$ whose price process $\{ S(t) \}_{t \ge 0}$ satisfies the following stochastic differential equation (SDE): 
\begin{equation}\label{ssde} 
\mathrm{d}S(t) = S(t_{-}) \bigl(\mu_{S}\,\mathrm{d}t + \sigma_{S}\, \mathrm{d}B_1(t) - \eta\,(\mathrm{d}N_1(t)-\lambda_{1}\, \mathrm{d}t)\bigr), \quad t \ge 0,
\end{equation} 
where $\mu_{S}$ is the (constant) expected rate of return; $\sigma_{S}$ is the constant volatility; $\lambda_1$ is the constant rate parameter of the Poisson process $N_1$; $\eta$ is the jump coefficient, which is supposed to be a positive constant. 
Consequently, a jump in the Poisson process $N_1$ causes a decrease in the price of the risky asset. 
Here, $S(t_{-})$, represents the price of the risky asset just before time $t$. 
The other financial asset is a risk-free security such as a money market account which pays a continuously compounded constant interest rate $r$. 
We assume as in \citet{damgaard2003} that the unit price of a durable good $P(t)$ follows a geometric Brownian motion as follows: 
\begin{equation}\label{psde}
\mathrm{d}P(t) = P(t) \bigl(\mu_{P}\, \mathrm{d}t + \sigma_{1}\, \mathrm{d}B_{1}(t) + \sigma_{2}\, \mathrm{d}B_2(t)\bigr), \quad t \ge 0,
\end{equation} 
where $\mu_{P}$, $\sigma_{1}$ and $\sigma_{2} \neq 0$ are supposed to be constants. 
We note that the unit price of the durable good is partially correlated with the price of the risky asset when $\sigma_{1}\neq0$. 

As in \citet{damgaard2003}, we assume that the stock of durable consumption goods depreciates at a constant rate $\delta^\dagger > 0$ over time. 
We further suppose that the durable consumption goods can be damaged by an insurable event, such as natural disasters, represented by the Poisson process $N_{2}$. 
We assume that a constant fraction $\ell \in (0,1)$ of the durable good is lost or damaged when an insured event occurs. 
This means that $K(t)$, the number of units of durable goods held by the agent at time $t$, changes due to natural causes and follows the SDE: 
\begin{equation}\label{ksde} 
\mathrm{d}K(t) = K(t_{-})\bigl(-\delta^{\dagger}\, \mathrm{d}t -\ell\, (\mathrm{d}N_{2}(t) - \lambda_2\, \mathrm{d}t) \bigr), 
\end{equation} 
in periods where the agent does not trade durable goods. 
For notational convenience and easier interpretability, we define $\delta \coloneqq \delta^{\dagger} - \ell \lambda_2$. 
Then Equation~\eqref{ksde} can be written as 
$\mathrm{d}K(t) = K(t_{-})\bigl(- \delta \mathrm{d}t -\ell \mathrm{d}N_{2}(t)\bigr)$. 
Here, we impose the condition $K(t) > 0$ due to the assumption that the economic agent cannot take a short position on the durable goods and to preclude the degenerate case that the agent does not invest in the durable goods.

We treat $K(t)$ as a state variable whose dynamics are given by~\eqref{ksde}. 
This is in line with specification in \citet{damgaard2003} and \citet{matos2014}. 
In the no-transaction-cost setting considered here, the agent can adjust the stock of durable goods continuously without frictions, which implies that the choice of $K(0)$ is effectively embedded in the control policy. 
The optimal strategy can therefore be represented in terms of proportional holdings $\alpha_k$ in~\eqref{vfunntc} below, and the functional $J(x,p; \Pi)$ below depends only on the initial wealth $x$ and durable-good price $p$. 

The economic agent may choose to purchase an insurance contract to cover the risk of loss. 
We denote by $q(t)\ge0$ the level of insurance coverage, i.e.\ the indemnity paid by the insurer at loss event time. 
The agent may freely choose or change the coverage level $q(t)$ at any time $t$ for the immediate future. 
We do not impose the assumption that $q(t)\le\ell K(t_-)P(t)$. 
That is, the agent is not prohibited from purchasing insurance coverage that exceeds the value of the durable good that is lost or damaged when the insured event occurs. 
We assume that the insurance premium is paid continuously over time and is given by
\begin{equation} 
I(t)=\varphi\lambda_{2}q(t),
\end{equation} 
subject to a loading factor $\varphi\ge1$. 

The economic agent must choose a consumption strategy and a trading strategy for the durable good and the financial assets. 
Let $c(t)$ be the consumption rate of perishable consumption goods at time $t$. 
We assume that the process $(c (t))_{t \ge 0}$ is $\mathbb{F}$-progressively measurable and satisfies certain integrability conditions. 
To meet her investment goals, the economic agent chooses a portfolio consisting of a risk-free asset, a risky asset, and a durable consumption good. 
Let $\pi_{0}(t)$ and $\pi_{1}(t)$ be the amounts invested in the risk-free asset and the risky asset at time $t$, respectively. 
The total wealth is given by the sum of the agent's investments in the risk-free and risky assets as well as the durable consumption good. 
Consequently, a trading strategy to be adopted by the agent consists of an $\mathbb{F}$-progressively measurable, integrable stochastic process $\bigl(\pi_{0}(t),\pi_{1}(t),K(t),q(t)\bigr)_{t \ge 0}$ with state space $\Re^2\times \Re_{>0}\times\Re_{\ge 0}$ such that the process satisfies certain integrability conditions. 
Note that $\Re_{>0}$ and $\Re_{\ge 0}$ denote the set of positive real numbers and the set of non-negative real numbers, respectively. 
Then the total wealth of the agent $X(t)$ at time $t$ is given by: 
\begin{equation}\label{totw}
X(t) = \pi_{0}(t) + \pi_{1}(t) + K(t)P(t), \quad t\ge0.
\end{equation} 

Assume that the agent follows a perishable consumption strategy $(c(t))_{t \ge 0}$ and a self-financing investment-insurance strategy $\bigl(\pi_{0}(t),\pi_{1}(t),K(t),q(t)\bigr)_{t \ge 0}$. 
Then the law of motion of the wealth process $(X(t))_{t \ge 0}$ is governed by the following stochastic differential equation: 
\begin{align}\label{dxt} 
\begin{aligned} 
\mathrm{d}X(t) & = rX(t)\, \mathrm{d}t - \bigl(c(t) + \varphi\lambda_{2}q(t)\bigr)\, \mathrm{d}t \\
& \quad + \bigl(\pi_{1}(t)(\mu_{S}+\eta\lambda_{1}-r) + K(t)P(t) (\mu_{P}-r-\delta)\bigr)\, \mathrm{d}t \\
& \quad + \bigl(\sigma_{S}\pi_{1}(t) + \sigma_{1}K(t)P(t)\bigr)\, \mathrm{d}B_{1}(t) + \sigma_{2}K(t)P(t)\, \mathrm{d}B_{2}(t) \\
& \quad - \eta\pi_{1}(t_{-})\, \mathrm{d}N_{1}(t) - \bigl(\ell K(t_{-})P(t) - q(t)\bigr)\, \mathrm{d}N_{2}(t), \quad t \ge 0.
\end{aligned} 
\end{align} 
By assumption, the total wealth of the agent satisfies the self-financing condition. 
That is, the changes in total wealth are due to fluctuations in the prices of the risky asset and the durable good, the natural depreciation of the durable good \eqref{ksde}, as well as the consumption of the perishable good. 
In particular, \eqref{dxt} holds even though the agent may continuously adjust positions of the assets, durable good, insurance or consumption. 

It is required that the agent's strategies satisfy the solvency condition. 
That is, at any moment, the economic agent's total wealth is always positive taking into account that the risky asset price process may jump and that an insured event may occur. 
These considerations lead to the solvency condition: 
\begin{align}\label{solvency1} 
\begin{aligned} 
& X(t_{-}) - \eta\pi_{1}(t_{-}) > 0, \\ 
& X(t_{-}) - \bigl(\ell K(t_{-})P(t) - q(t)\bigr) > 0, 
\end{aligned} 
\end{align} 
$\mathbb{P}$-almost surely for each $t>0$. 
Condition~\eqref{solvency1} means that the agent's total wealth must be able to withstand a sudden crash of the risky asset price, or a sudden damage to the durable good due to an insurable event, after the indemnity has been paid. 

Let $(x,k,p)$ denote the initial values of the state variables $(X(0),K(0),P(0))$. 
The initial values of the state variables $(x,k,p)$ are said to be in the solvency region if $x>0$, $k>0$ and $p>0$. 
Recall the assumption that $q (t) \ge 0$ and $K (t) > 0$, for all $t \ge 0$. 

A strategy $\Pi \coloneqq (\Pi (t))_{t \ge 0}$, where  $\Pi(t)=(c(t),\pi_{0}(t),\pi_{1}(t),K(t),q(t))$ takes values in
$\Re_{>0}\times\Re\times\Re\times\Re_{>0}\times\Re_{\ge0}$, is admissible if the corresponding wealth process $(X(t))_{t \ge 0}$ satisfies the solvency condition~\eqref{solvency1} and $X(t)>0$ for all $t\ge0$, almost surely. 
We denote by $\mathcal{A}(x,p)$ the set of admissible strategies for initial state $(x,p)$. 
For all $(x,k,p)$ in the solvency region, admissible strategies exist. 
For example, fix some $\widetilde{K}\in (0, 1/\ell)$ and let the agent hold $K(t)\equiv \widetilde{K}X(t)/P(t)$ for all $t\ge 0$, invest all remaining wealth in the risk-free asset, choose 
$c(t)\equiv \widetilde{c}X(t)$ with $\widetilde{c}>0$ small enough, and set $q(t)\equiv 0$. 
This produces a strictly positive wealth process $X$ and a durable stock $K$ that remains strictly positive for all $t\ge 0$. 
Hence the set of admissible strategies is non-empty.

Following \cite{damgaard2003} and \citet{matos2014}, we consider an economic agent who maximizes a time-separable utility function on perishable and durable goods consumption over an infinite time horizon, given by
\begin{equation}\label{UTL}
\int_{0}^{\infty}e^{-\rho t}u(c(t),K(t))\, \mathrm{d}t,
\end{equation} 
where the constant $\rho > 0$ is the subjective discount rate of the economic agent.

We assume that the utility of the consumption function exhibits a constant relative risk aversion, given by
\begin{equation}\label{ut}
u(c,k)=\frac{1}{1-\gamma}\bigl(c^{\beta}k^{1-\beta}\bigr)^{1-\gamma}, \quad \beta \in (0,1),\; \gamma> 0, 
\end{equation} 
where $\gamma$ is the constant of risk aversion and the constant $\beta$ determines the relative weight of perishable and durable goods. 
In the limiting case $\gamma=1$, the utility reduces to $u(c,k)=\ln(c^{\beta}k^{1-\beta})$. Note that \cite{damgaard2003} and \cite{matos2014} consider $0<\gamma<1$ only.
The agent's objective is to find an admissible strategy $\Pi$ such that the expected utility at $t=0$, given by 
\begin{equation}\label{J}
J(x,k,p;\Pi) = \mathbb{E}\Bigl[\int_{0}^{\infty}e^{-\rho t}u(c(t),K(t))\, \mathrm{d}t\Bigr]
\end{equation} 
is maximized, where $X(0)=x$, $K(0)=k$ and $P(0)=p$. 
Note that the expected utility is independent of time, because we consider an infinite time horizon and time-homogeneous state equations \eqref{ssde}--\eqref{ksde}. 
Moreover, the expected utility depends only on $x$ and $p$, since in the absence of transaction costs $K(0)$ can be instantaneously adjusted to the level indicated by $\Pi$ without any cost. 
The value function of the agent is thus given by: 
\begin{equation}\label{V}
V(x,p) = \sup_{\Pi \in \mathcal{A}(x,p)} J(x,p;\Pi). 
\end{equation} 
By the principle of dynamic programming, the value function satisfies 
\begin{equation}\label{dpv}
V(x,p) = \sup_{\Pi \in \mathcal{A}(x,p)} \mathbb{E}\Bigl[\int_{0}^{t} e^{-\rho s}u(c(s),K(s))\, \mathrm{d}s +e^{-\rho t}V(X(t),P(t))\Bigr], 
\end{equation} 
for all $t \ge 0$, where $X(0)=x$ and $P(0)=p$. 

\section{Solution and verification theorem}\label{notransactioncos_sec}
In this section, we study a slightly extended version of the optimal control problem~\eqref{V} using the Hamilton--Jacobi--Bellman (HJB) dynamic programming approach.
A verification theorem will be provided along the lines of Theorem 2.2 in \cite{framstadINRIA2001optimal}, (see also Theorem 1 in \cite{framstad2001optimal}). 
Then as in Theorem 2.3 in \cite{framstadINRIA2001optimal}, it will be proved that a candidate value function and the respective candidate optimal control are the optimal solution to the slight extension of the optimal control problem in \eqref{V}.

First, we shall present the HJB equation and some related calculations to find the candidate value function and the candidate optimal control processes for the optimal control problem in \eqref{V}.
Then the slight extension of the problem in \eqref{V} is presented for which the verification theorem will be proved. Some auxiliary results are presented in lemmas. A condition in the verification theorem is related
to a generalization of the transversality condition in \citet{damgaard2003} by incorporating jumps.
The proofs of all the lemmas and theorems will be provided in Appendices A, B and C. 

\subsection{Semi-explicit solution of HJB equation}
Applying Ito's formula for jump-diffusion processes, the Hamilton--Jacobi--Bellman (HJB) equation corresponding to (\ref{dpv}) is obtained as: 

\beqr{hjb0}
0 &= \underset{(c,\pi_1,k,q)}{\sup}\Big\{ \frac{1}{1-\gamma}(c^{\beta}k^{1-\beta})^{1-\gamma} - \rho V(x,p)\\
&+\big(rx+\pi_1(\mu_{S}+\lmb_1\eta-r)+kp(\mu_{P}-r-\delta)-\varphi\lmb_2q-c\big)\frac{\pd V(x,p)}{\pd x} \\
&+\frac{1}{2}\big(\pi_1^{2}\sigma_{S}^{2}+k^{2}p^{2}(\sigma_{1}^{2}+\sigma_{2}^{2})+2\pi_1kp\,\sigma_{S}\sigma_{1}\big)\frac{\pd^{2}V(x,p)}{\pd x^{2}} \\
& +\mu_{P}p\frac{\pd V(x,p)}{\pd p}+\frac{1}{2}(\sigma_{1}^{2}+\sigma_{2}^{2})p^{2}\frac{\pd^{2}V(x,p)}{\pd p^{2}} \\
&+\big(kp^2(\sigma_{1}^2+\sigma_{2}^2)+\pi_1p\,\sigma_{1}\sigma_{S}\big)\frac{\pd^{2}V(x,p)}{\pd x\pd p} \\
& +\lmb_1\big(V(x-\eta\pi_1,p)-V(x,p)\big) \\
& +\lmb_2\big(V(x-(\ell kp-q),p)-V(x,p)\big)\Big\}.
\eeqr

We note that (\ref{hjb0}) is time-independent. Then it will be shown that the (candidate) optimal strategies are constant-fraction strategies, which are proportional to the current wealth or the ratio of the
current wealth and the current unit price of the durable good. Following \citet{damgaard2003} and \citet{matos2014} arguments
and using the homogeneity of the instantaneous utility
function to reduce the dimensionality of the problem, we note that
for $\kappa>0,\,\big(c,\pi_1,k,q\big)$ is admissible with initial
wealth and initial durable price $(x,p)$ if and only if $(\kappa c, \kappa\pi_1,k,\kappa q)$
is admissible with initial wealth and initial durable price $(\kappa x,\kappa p)$.
Since $u(\kappa c,k)=\kappa^{\beta(1-\gamma)}u(c,k)$,
it implies that $V\big(\kappa x,\kappa p\big)=\kappa^{\beta(1-\gamma)}V\big(x,p\big)$
and in particular $V\big(\frac{x}{p},1\big)=p^{-\beta(1-\gamma)}V\big(x,p\big)$ for $\kappa=\frac{1}{p}$. Defining the reduced value function $v(y)=v(\frac{x}{p})=V(\frac{x}{p},1)$ and substituting into  (\ref{hjb0}), we obtain the HJB equation for $v$ as
\beqr{hjby}
0 =& \underset{(\hat{c}, \hat{\pi}_1,k,\hat{q})}{\sup}\bigg\{
\frac{1}{1-\gamma}\big(\hat{c}^{\beta}k^{1-\beta}\big)^{1-\gamma} \\
&
+v(y)\Big(\beta(1-\gamma)\mu_P-\rho-\frac{1}{2}\sigma_P^2\beta(1-\gamma)(1-\beta(1-\gamma))\Big) \\
& +v'(y)\Big((r+(1-\beta(1-\gamma))\sigma_P^2-\mu_{P})(y-k)-\delta k-\varphi\lmb_2\qh \\
&
+\hat{\pi}_1(\mu_{S}+\lmb_1\eta-r-(1-\beta(1-\gamma))\sigma_{S}\sigma_{1})-\hat{c}\Big) \\
& +v''(y)\Big(\frac{1}{2}\hat{\pi}_1^{2}\sigma_{S}^{2}+\frac{1}{2}\sigma_P^2(y-k)^{2}-\hat{\pi}_1\sigma_{S}\sigma_{1}(y-k)\Big) \\
&
+\lmb_1(v(y-\eta\hat{\pi}_1)-v(y))
+\lmb_2(v(y-(\ell k-\qh))-v(y))\bigg\},
\eeqr
where we let $\sigma_P^2=\sigma_{1}^2+\sigma_{2}^2$. Here, $v',v''$ are the first and second derivatives
of $v$, and the normalized controls $\hat{c}=c/p$, $\hat{\pi}_1=\pi_1/p$ and $\qh=q/p$. We claim that (\ref{hjby}) has a solution of the form
$v(y)=\frac{1}{1-\gamma}\ap_{v}y^{1-\gamma}$ with optimal
control variables $\hat{c}=\ap_{c}y,\,\hat{\pi}_1=\alpha_{\pi}y,\,k=\ap_{k}y$
and $\hat{q}=\ap_{q}y$. Here, $\ap_{k},\ap_{c},\ap_{q}$ represent the optimal holding position of the durable consumption good, the optimal rate of consumption for the perishable good and the optimal level of the insurance, all in terms of fractions of total wealth. Substituting these fractions into (\ref{hjby}), we obtain
\beqr{hjba}
0 =& \underset{ (\ap_c, \alpha_{\pi},\ap_k,\ap_{q})}{\sup}\bigg\{
\frac{1}{1-\gamma}\big(\ap_{c}^{\beta}\ap_k^{1-\beta}\big)^{1-\gamma} \\
&
+\frac{\ap_v}{1-\gamma}\Big(\beta(1-\gamma)\mu_P-\rho-\frac{1}{2}\sigma_{P}^{2}\beta(1-\gamma)(1-\beta(1-\gamma))\Big) \\
& +\ap_v\Big((r+(1-\beta(1-\gamma))\sigma_{P}^{2}-\mu_{P})(1-\ap_k)-\delta\ap_k-\varphi\lmb_2\ap_q \\
&
+\alpha_{\pi}(\mu_{S}+\lmb_1\eta-r-(1-\beta(1-\gamma))\sigma_{S}\sigma_{1})-\ap_{c}\Big) \\
& - \gamma\ap_v\Big(\frac{1}{2}\alpha_{\pi}^{2}\sigma_{S}^{2}+\frac{1}{2}\sigma_{P}^{2}(1-\ap_k)^{2}-\alpha_{\pi}\sigma_{S}\sigma_{1}(1-\ap_k)\Big) \\
&
+\frac{\ap_v}{1-\gamma}
\Big(\lmb_1((1-\eta\alpha_{\pi})^{1-\gamma}-1)
+\lmb_2((1-\ell\ap_k+\ap_q)^{1-\gamma}-1)\Big)\bigg\} \\
=:& \underset{ (\ap_c, \alpha_{\pi},\ap_k,\ap_{q})}{\sup}
H (\ap_c, \alpha_{\pi},\ap_k,\ap_{q}), \quad \mbox{say}.
\eeqr
Calculation of supremum in the above is subject to the solvency conditions (\ref{solvency1}) and model constraints: \begin{equation}\label{param_constraints}
1-\eta\alpha_{\pi} > 0,\;	1-\ell\alpha_k+\alpha_q > 0,\;  \alpha_c>0,\;\alpha_k> 0,\;\alpha_q\ge 0.
\end{equation}
The constraints in~\eqref{param_constraints} are the natural portfolio-share counterparts of the solvency condition~\eqref{solvency1}: they ensure that jumps in the risky asset and in the durable good do not drive the wealth process to zero or negative values, which is essential for the subsequent positivity and martingale arguments in Lemma~\ref{nobankruptcy}. 

It is possible to prove that $H (\ap_c, \alpha_{\pi},\ap_k,\ap_{q})$ is a concave function (for a proof, see  \ref{ConcavityProof_sec})\footnote{Note that only first order conditions are considered in models studied in \cite{damgaard2003} and \cite{matos2014} without proof of concavity. Our model is generalisation of these models and thus our proof can apply to these special cases too.} and thus the \emph{Karush--Kuhn--Tucker conditions} are sufficient conditions to be satisfied at supremum of $H$; see \citet[Section 5.5.3, p.244]{boyd2004convex}, i.e.\ the supremum is achieved either at the first order conditions or at the boundary. 
After some algebra, see \ref{ConcavityProof_sec}, the Karush--Kuhn--Tucker conditions can be written as the following system of equations
when $\sigma_{1}\neq0$ 
\beqc{optc}
\ap_{v}=  \beta\ap_{c}^{\beta(1-\gamma)-1}\ap_{k}^{(1-\beta)(1-\gamma)},\\
\ap_{k}=  \frac{1}{\gamma\sigma_{S}\sigma_{1}}\big(\mu_{S}+\lmb_1\eta-r -\gamma\alpha_{\pi}\sigma_{S}^{2}-(1-\beta)(1-\gamma)\sigma_{S}\sigma_{1}-\eta\lmb_1(1-\eta\alpha_{\pi})^{-\gamma}\big), \\
\ap_{c}= \frac{\beta}{1-\beta}\big(\sigma_{P}^{2}((1-\beta)(1-\gamma)+\gamma\ap_k)+\gamma\alpha_{\pi}\sigma_{S}\sigma_{1}+r-\mu_{P}+\delta\\
\qquad +\ell\lmb_2(1-\ell\ap_{k}+\ap_{q})^{-\gamma}\big)\ap_{k}, \\
\ap_{q}=  \max(\ell\ap_{k}-1+\varphi^{-\frac{1}{\gamma}},0),
\eeqc
and when $\sigma_{1}=0$ (i.e. zero correlation between durable good unit price and risky asset processes)
\beqc{optc0}
	\ap_{v}=  \beta\ap_{c}^{\beta(1-\gamma)-1}\ap_{k}^{(1-\beta)(1-\gamma)},\\
	0 =  \mu_{S}+\lmb_1\eta-r -\gamma\alpha_{\pi}\sigma_{S}^{2}-\eta\lmb_1(1-\eta\alpha_{\pi})^{-\gamma}, \\
	\ap_{c}=\frac{\beta}{1-\beta}\big(\sigma_{P}^{2}((1-\beta)(1-\gamma)+\gamma\ap_k)+r-\mu_{P}+\delta+\ell\lmb_2(1-\ell\ap_{k}+\ap_{q})^{-\gamma}\big)\ap_{k}, \\
	\ap_{q}=  \max(\ell\ap_{k}-1+\varphi^{-\frac{1}{\gamma}},0).
\eeqc
Substituting the optimal controls given by (\ref{optc}) or (\ref{optc0}) into (\ref{hjba}),
we numerically solve for the optimal control $\alpha_{\pi}$  or $\ap_k$, respectively, (subjecting to parameter constraints (\ref{param_constraints})), followed by other controls by substitution.
After numerically solving for the constants $(\ap_c, \alpha_{\pi},\ap_k,\ap_{q})$, we obtain the (candidate) value function:
\beql{vfunntc}
V(x,p)=\frac{1}{1-\gamma}\ap_{v}p^{-(1-\beta)(1-\gamma)}x^{(1-\gamma)},
\eeq
and the respective (candidate) optimal control processes are given as follows:
\beqr{optimalcontrol}
\pi_1(t)=\alpha_{\pi}X(t), \ K(t)=\ap_{k}\frac{X(t)}{P(t)}, \ c(t)=\ap_{c}X(t), \ q(t)=\ap_{q}X(t), 
\eeqr
where $X(t)$ is the wealth process governed by (\ref{dxt})
corresponding to the (candidate) optimal control
processes in (\ref{optimalcontrol}). Consequently,
the (candidate) optimal control processes
are feedback controls. The coefficients of the (candidate) optimal control processes in (\ref{optimalcontrol}) satisfy (\ref{optc}) when $\sigma_{1}\neq0$, and they satisfy (\ref{optc0}) when $\sigma_{1}=0$. The remaining wealth is then invested in the risk-free account.

\subsection{Verification theorem}

In this subsection, we shall first provide a verification theorem along the line of Theorem 2.2 in Framstad et al. (2001a), (see also Theorem 1 in Framstad et al. (2001b)). Then as in Theorem 2.3 in Framstad et al. (2001a), we prove the candidate value function in (\ref{vfunntc}) and the candidate optimal control processes in (\ref{optimalcontrol}) are the solution of the optimal control problem in (\ref{V}) if a certain transversality condition is satisfied. 

Let $A = -(1-\beta)(1-\gamma)$ and $B = 1-\gamma$.
Then the candidate value function in (\ref{vfunntc})
is written as follows:
\beql{vfunntc1}
V(x,p)=\frac{1}{1-\gamma}\ap_{v}p^A x^B.
\eeq
Let ${\cal O} = (0, \infty) \times (0, \infty)$, where ${\cal O}$ is the solvency region. For each $(s, x, p) \in \Re_{\ge 0} \times {\cal O}$ and each admissible strategy $\Pi$, the following performance
functional is considered:
\beql{GJ}
J(s,x,p,{\Pi})=\mathbb{E}^{(s, x, p)} \Prt{\int_{0}^{\infty} e^{-\rho (s + t)}u(c(t),K(t))dt},
\eeq 
where $\mathbb{E}^{(s, x, p)}$ is the expectation taken under
the probability measure $\mathbb{P}^{(s, x, p)}$; $\mathbb{P}^{(s, x, p)}$ represents the probability laws of the joint process
$\{ (s + t, X(t), P(t)) \}_{t \ge 0}$ given that $(s, X(0), P(0)) = (s, x, p) \in \Re_{\ge 0} \times {\cal O}$ under the probability measure $\mathbb{P}$. 

When $s = 0$, the performance functional in (\ref{GJ})
reduces to:
\beql{GJ0}
J(0,x,p,{\Pi})=\mathbb{E}^{(0, x, p)} \Prt{\int_{0}^{\infty} e^{-\rho t}u(c(t),K(t))dt}.
\eeq 
Consequently, the performance functional in (\ref{GJ0}) coincides
with that in (\ref{J}). 

Using the performance functional in (\ref{GJ}), the value
function of the slight
extension to the optimal control problem in (\ref{V}) 
is given by:
\beql{GV}
V(s,x,p)=\underset{{\Pi}}{\sup}J(s,x,p,{\Pi}).
\eeq
Then
\beql{GV1}
V(s,x,p) = \underset{{\Pi}}{\sup}J(s,x,p,{\Pi}) = \underset{{\Pi}}{\sup} e^{-\rho s} J(0,x,p,{\Pi}) = 
e^{-\rho s} \underset{{\Pi}}{\sup} J(x,p,{\Pi})
= e^{-\rho s} V(x, p).
\eeq
From (\ref{vfunntc}) and (\ref{GV1}), we conjecture that
the candidate value function of the optimal control
problem in (\ref{GV}) is given by:
\beql{gvfunntc}
V(s,x,p)=\frac{1}{1-\gamma}\ap_{v} e^{-\rho s} p^A x^B.
\eeq
From (\ref{GV1}), we note that the respective optimal control processes remain the same as those in (\ref{optimalcontrol}).

Define the vector-valued process $\{ R (t) \}_{t \ge 0}$
by $R (t) = (s + t, X(t), P(t))^\top$ such that 
$R (0) = (s, X(0), P(0)) = (s, x, p) \in \Re_{\ge 0} \times {\cal O}$, say, where 
${\bf y}^\top$ is the transpose of a
vector $\bf y$. Recall $\Re_{\ge 0}$ is the set of non-negative real numbers. This type of vector-valued processes
was considered in \cite{oksendal2019applied}.

Let ${\cal A}^{(c, \pi_1, k, q)}_R$ denote the infinitesimal generator of the vector-valued process $\{ R (t) \}_{t \ge 0}$ corresponding to a given 
admissible control $(c, \pi_1, k, q) := (c (y), \pi_1 (y), k (y), q (y))$, where $y = \frac{x}{p} \ge 0$, for any $(s, x, p) \in \Re_{\ge 0} \times {\cal O}$. Let ${\cal C}^{1, 2} (\Re_{\ge 0} \times {\cal O})$ denote the space of functions $\phi (s, x, p)$ on $\Re_{\ge 0} \times {\cal O}$, which are continuously differentiable in $s \in \Re_{\ge 0}$ and twice continuously differentiable in $(x, p) \in {\cal O}$. Then for any $\phi (s, x, p) \in {\cal C}^{1, 2} (\Re_{\ge 0} \times {\cal O})$,
\begin{eqnarray} \label{ARgenerator}
&& {\cal A}^{(c, \pi_1, k, q)}_R [\phi (s, x, p)] \nonumber\\
&=& \frac{\partial \phi}{\partial s} + \big(rx+\pi_1(\mu_{S}+\lmb_1\eta-r)+kp(\mu_{P}-r-\delta)-\varphi\lmb_2q-c\big)\frac{\pd \phi}{\pd x} \nonumber\\
&&+\frac{1}{2}\big(\pi_1^{2}\sigma_{S}^{2}+k^{2}p^{2}(\sigma_{1}^{2}+\sigma_{2}^{2})+2\pi_1kp\,\sigma_{S}\sigma_{1}\big)\frac{\pd^{2} \phi}{\pd x^{2}} \nonumber\\
&& +\mu_{P}p\frac{\pd \phi}{\pd p}+\frac{1}{2}(\sigma_{1}^{2}+\sigma_{2}^{2})p^{2}\frac{\pd^{2} \phi}{\pd p^{2}} +\big(kp^2(\sigma_{1}^2+\sigma_{2}^2)+\pi_1p\,\sigma_{1}\sigma_{S}\big)\frac{\pd^{2}\phi}{\pd x\pd p} \nonumber\\
&& +\lmb_1\big(\phi(x-\eta\pi_1,p)-\phi(x,p)\big) +\lmb_2\big(\phi(x-(\ell kp-q),p)-\phi(x,p)\big).
\end{eqnarray}

The following lemma presents an auxiliary result which is related to a no-bankruptcy condition. It will be used in the proof of Lemma \ref{Integrability} in Appendix A. It will also be used in the proof of Theorem \ref{verificationCand2} in Appendix A.


\begin{lem} \label{nobankruptcy}
Let $\{ X(t) \}_{t \ge 0}$ be the wealth process governed by \eqref{dxt} under the feedback controls \eqref{optimalcontrol}. Assume that the constants $(\alpha_c, \alpha_\pi, \alpha_k, \alpha_q)$ satisfy the constraints in \eqref{param_constraints}. 
If $X(0) > 0$, then 
\begin{equation} \label{PositiveX(t)}
X(t) > 0, \qquad \forall t \ge 0, \ \mathbb{P}\text{-a.s.} 
\end{equation} 
Furthermore, Eq. (\ref{PositiveX(t)}) implies that for any $\theta \in \mathbb{R}$, $X(t)^\theta$ is well defined and positive, for all $t \ge 0$, $\mathbb{P}$-a.s. 
\end{lem} 
Note that Eq. (\ref{PositiveX(t)}) in Lemma \ref{nobankruptcy} may be interpreted as the $\mathbb{P}$-a.s. no-bankruptcy condition. \\ 

\noindent
The following lemma gives an auxiliary result in the form of an integrability condition for the candidate value function 
$V (s, x, p)$ in (\ref{gvfunntc}). This lemma will be used in the proof of Theorem \ref{verificationCand1} in Appendix A to justify that the candidate value function $V (s, x, p)$ in (\ref{gvfunntc}) satisfies Condition (a)(ii) in Theorem \ref{verification}. 

\begin{lem} \label{Integrability}
Let 
\begin{eqnarray} \label{T1}
T_1 &=& B (r + \ap_{\pi} (\mu_S + \lambda_1 \eta - r) + \ap_k
(\mu_P - r - \delta) - \varphi \lambda_2 \ap_q - \ap_c) \nonumber\\
&& + 
\frac{1}{2}B(B - 1) ((\ap_{\pi} \sigma_S + \ap_k \sigma_1)^2 + \ap^2_k \sigma^2_2) \nonumber\\
&& + ((1 - \ap_{\pi} \eta)^{B} - 1) \lambda_1 + ((1 - \ap_k l + \ap_q)^{B} - 1) \lambda_2 \nonumber\\
&& + A \bigg (\mu_P + \frac{1}{2} (A - 1) (\sigma^2_1 + \sigma^2_2) \bigg ) + AB(\ap_{\pi} \sigma_S \sigma_1 + \ap_k (\sigma^2_1 + \sigma^2_2)).
\end{eqnarray}
Suppose that $\rho > T_1$ and that the control is given by the feedback control $(c, \pi_1, k, q)$ in~\eqref{optimalcontrol}. 
Then, for any finite stopping time $\tau$, 
\beqr{Condition3}
& \mathbb{E}^{(s, x, p)} \bigg [ | V (R(\tau)) | + \int^{\tau}_0 
| {\cal A}^{(c, \pi_1, k, q)}_R [V(R(t)] | dt \bigg ] < \infty,
\eeqr
where ${\cal A}^{(c, \pi_1, k, q)}_R$ is the infinitesimal generator given by (\ref{ARgenerator}); $V (R(t)) = V(s+t, X (t), P(t))$, which 
is given by (\ref{gvfunntc}).
\end{lem}

Note that the condition that $\rho > T_1$ in Lemma
\ref{Integrability} is a generalization of the transversality condition in \cite{damgaard2003} to incorporate jumps. 

The following theorem gives the verification theorem for the solution of the optimal control problem in (\ref{GV}). The verification theorem is related to Theorem 2.2 in \cite{framstadINRIA2001optimal}, (see also Theorem 1 in \cite{framstad2001optimal}).

\sloppy
\begin{thm} \label{verification}
(a) Suppose there exists a non-negative function $\nu (s, x, p) 
\in {\cal C}^{1, 2} (\Re_{\ge 0} \times {\cal O})$ such that 
\begin{enumerate}
\item[(i)] the following inequality holds:
\begin{eqnarray} \label{Cond(a)(i)}
{\cal A}^{(c, \pi_1, k, q)}_R [\nu (s, x, p)] + u (c, k) \le 0, 
\end{eqnarray}
for any admissible control $(c, \pi_1, k, q) \in \Re_{> 0} \times \Re \times \Re_{> 0} \times \Re_{\ge 0}$.
\item[(ii)] $\nu (s, x, p)$ satisfies the condition in (\ref{Condition3}). That is,
\beqr{Condition3nu}
& \mathbb{E}^{(s, x, p)} \bigg [ | \nu (R(\tau)) | + \int^{\tau}_0 
| {\cal A}^{(c, \pi_1, k, q)}_R [ \nu (R(t)] | dt \bigg ] < \infty,
\eeqr
for any admissible control $(c, \pi_1, k, q) \in \Re_{> 0} \times \Re \times \Re_{> 0} \times \Re_{\ge 0}$, where $\nu (R(t)) = \nu (s+t, X(t), P(t))$.
\end{enumerate}
Then 
\begin{eqnarray} \label{Conclusion(a)}
\nu (s, x, p) \ge V (s, x, p).
\end{eqnarray}
(b) 
Suppose, in addition to Conditions (a)(i) and (a)(ii) of this theorem, that there exists
$({\hat c}, {\hat \pi}_1, {\hat k}, {\hat q}) := ({\hat c} (y), {\hat \pi}_1 (y), {\hat k} (y), {\hat q} (y)) \in \Re_{> 0} \times \Re \times \Re_{> 0} \times \Re_{\ge 0}$ such that
\begin{enumerate}
\item[(i)] 
\begin{eqnarray} \label{Condition(b)(i)}
{\cal A}^{({\hat c} (y), {\hat \pi}_1 (y), {\hat k} (y), {\hat q} (y))}_R [\nu (s, x, p)] + u ({\hat c} (y), {\hat k} (y)) = 0. 
\end{eqnarray}
\item[(ii)] $({\hat c}, {\hat \pi}_1, {\hat k}, {\hat q}) := ({\hat c} (y), {\hat \pi}_1 (y), {\hat k} (y), {\hat q} (y))$ is an admissible strategy.
\item[(iii)] The following transversality condition holds:
\begin{eqnarray} \label{Condition(b)(iii)}
\lim_{L \to \infty} \mathbb{E}^{(s, x, p)} \left [e^{- \rho T_L} \nu (R (T_L)) \right ] = 0, \quad \mbox{for all $(s, x, p) \in \Re_{\ge 0} \times {\cal O}$}.
\end{eqnarray}
Note that $T_L := \min (L, \inf \{ t > 0; | {\hat X} (t) | \ge L \} )$, where $L \in (0, \infty)$ and $\{ {\hat X} (t) \}_{\ge 0}$ is the wealth process evaluated at the Markov control $({\hat c}, {\hat \pi}_1, {\hat k}, {\hat q}) := ({\hat c} (y), {\hat \pi}_1 (y), {\hat k} (y), {\hat q} (y))$.  
\end{enumerate}
Then 
\begin{eqnarray} \label{Conclusion(b)}
\nu (s, x, p) = V (s, x, p),
\end{eqnarray}
and the Markov control $({\hat c}, {\hat \pi}_1, {\hat k}, {\hat q})$
is an optimal control of the problem in (\ref{GV}). 
\end{thm}

The following theorem proves that when $\gamma \in (0, 1)$, the candidate value function in (\ref{gvfunntc}) is the value function, and the candidate optimal control in (\ref{optimalcontrol}) is the respective optimal control. It is related to Theorem 2.3 in Framstad et al. (2001a).

\begin{thm} \label{verificationCand1}
Let $({\hat c}, {\hat \pi}_1, {\hat k}, {\hat q}) := ({\hat c} (y), {\hat \pi}_1 (y), {\hat k} (y), {\hat q} (y))$ such that
\begin{eqnarray} \label{MarkovControl}
{\hat c} (y) = \alpha_c y, \quad {\hat \pi}_1 (y) = \alpha_{\pi} y, \quad {\hat k} (y) = \alpha_k y, \quad {\hat q} (y) = \alpha_q y.
\end{eqnarray}
Suppose that
\begin{enumerate}
\item[(i)] $(\alpha_c, \alpha_{\pi}, \alpha_k, \alpha_q)$ satisfies (\ref{optc}) when $\sigma_{1} \neq 0$, and it satisfies (\ref{optc0}) when $\sigma_{1} = 0$.
\item[(ii)] $(\alpha_c, \alpha_{\pi}, \alpha_k, \alpha_q)$ satisfies the constraints in (\ref{param_constraints}).
\item[(iii)] $\rho > T_1$, where $T_1$ is given by (\ref{T1}).
\end{enumerate}
Then, when $\gamma \in (0, 1)$, the candidate value function $V (s, x, p)$ in (\ref{gvfunntc}) is the value function, and $({\hat c}, {\hat \pi}_1, {\hat k}, {\hat q})$ is the respective optimal control of the problem in (\ref{GV}).
\end{thm}

The following theorem shows that when $\gamma > 1$, the candidate value function in (\ref{gvfunntc}) is the value function, and the candidate optimal control in (\ref{optimalcontrol}) is the respective optimal control. Theorem 5.1 of \cite{oksendal2019applied}  will be used to prove this theorem.

\begin{thm} \label{verificationCand2}
Let 
\begin{eqnarray} \label{T1dagger}
T^{\dagger}_1 &=& 2 B (r + \ap_{\pi} (\mu_S + \lambda_1 \eta - r) + \ap_k
(\mu_P - r - \delta) - \varphi \lambda_2 \ap_q - \ap_c) \nonumber\\
&& + B(2B - 1) ((\ap_{\pi} \sigma_S + \ap_k \sigma_1)^2 + \ap^2_k \sigma^2_2) \nonumber\\
&& + ((1 - \ap_{\pi} \eta)^{2B} - 1) \lambda_1 + ((1 - \ap_k l + \ap_q)^{2B} - 1) \lambda_2 \nonumber\\
&& + 2 A \bigg (\mu_P + \frac{1}{2} (2 A - 1) (\sigma^2_1 + \sigma^2_2) \bigg ) + 4 AB(\ap_{\pi} \sigma_S \sigma_1 + \ap_k (\sigma^2_1 + \sigma^2_2)),
\end{eqnarray}
Suppose that the following two conditions hold: 
\begin{enumerate}
\item[(i)] $\rho > \frac{T^{\dagger}_1}{2}$, where $T^{\dagger}_1$ is given by (\ref{T1dagger}).
\item[(ii)] Conditions (i)--(iii) from Theorem \ref{verificationCand1} hold, (in particular, $\rho > T_1$, where $T_1$ is given by (\ref{T1})). 
\end{enumerate}
Then, when $\gamma > 1$, the candidate value function $V (s, x, p)$ in (\ref{gvfunntc}) is the value function, and $({\hat c}, {\hat \pi}_1, {\hat k}, {\hat q})$ is the respective optimal control of the problem in (\ref{GV}).
\end{thm}

{In summary, the optimal consumption and investment strategy of the economic agent consists of four components: the consumption of the perishable good $c(t),$ the value $P(t)K(t)$ invested in the durable good, the level of insurance coverage $q(t)$ on the durable
good, and the wealth $\pi_1(t)$ invested in the risky asset. All these components are constant fractions of the total wealth $X(t)$. These results extend those obtained by \citet{matos2014} to include a Poisson process modelling the insurance events and optional insurance coverage on the durable good. The optimal level of insurance coverage is directly affected by the loading factor $\varphi$, and is a decreasing function of the latter. Specifically, when insurance becomes too expensive, the agent will choose not to purchase any insurance. The optimal insurance coverage is also affected by the risk aversion coefficient $\gamma$: the greater the risk aversion, the higher the optimal insurance coverage. 

\section{Numerical experiments}
\label{numerical_sec}
%

To investigate optimal strategies obtained using our derived semi-explicit solution for optimal allocations (\ref{optimalcontrol}), in this section we present numerical examples.
We consider the baseline  parameter values presented in Table \ref{tab:parameters} aligned with the ones used in \cite{damgaard2003} and \cite{matos2014}.
In Figures \ref{sensitivity-loads}, \ref{sensitivity-frequency}, \ref{sensitivity-fraction} and \ref{sensitivity-aversion} we show the optimal wealth allocations into consumption, risky asset, durable good and insurance coverage level ($\ap_c$, $\alpha_{\pi}$, $\ap_k$, and $\ap_{q}$ respectively), calculated when one of the parameters is varying. 
Specifically, Figure \ref{sensitivity-loads} shows optimal allocations as a function of the insurance load $\varphi$. 
Figure \ref{sensitivity-frequency} shows optimal allocations when annual  frequency of durable good damage events  $\lambda_2$ is changing. 
Figure \ref{sensitivity-fraction} shows results for optimal allocations under different values for the damage fraction $\ell$. 
Finally, Figure~\ref{sensitivity-aversion} shows the sensitivity of the optimal allocations with respect to the risk aversion parameter. 

\begin{table}[htbp]
\small 
\centering
\begin{threeparttable}
\caption{Parameter settings used in the numerical experiments}
\label{tab:parameters}
\begin{tabular}{ccl}
\toprule
\textbf{Parameter} & \textbf{Value} & \textbf{Description} \\
\midrule
$r$ & 0.03 & Risk-free rate \\
$\rho$ & 0.03 & Subjective discount rate \\
$\gamma$ & 0.50 & Risk aversion coefficient \\
$\beta$ & 0.75 & Preference weight on perishable consumption \\
$\mu_S$ & 0.05 & Expected return of risky asset \\
$\sigma_S$ & 0.20 & Volatility of risky asset \\
$\lambda_1$ & 1.00 & Frequency of jump in risky asset price \\
$\eta$ & 0.05 & Severity of jump in risky asset price \\
$\mu_P$ & 0.08 & Drift of durable good price \\
$\sigma_{1}$ & 0.10 & Idiosyncratic volatility of durable good \\
$\sigma_{2}$ & 0.20 & Systematic volatility of durable good \\
$\delta^{\dagger}$ & 0.03 & Depreciation rate of durable good \\
$\lambda_2$ & 0.05 & Arrival rate of rare insurable events \\
$\ell$ & 0.50 & Loss proportion in insurable event \\
$\varphi$ & 1.10 & Insurance premium loading factor \\
\bottomrule
\end{tabular}
\end{threeparttable}
\end{table} 

The results are consistent with intuitive expectations. 
Figure~\ref{sensitivity-loads} shows that the demand for insurance and for the durable good is highly sensitive to the insurance loading, while perishable consumption is essentially unaffected. 
When the premium is actuarially fair $(\varphi = 1)$, the agent chooses the maximal insurance coverage and a relatively high allocation to the durable good. 
As the loading $\varphi$ increases, the optimal insurance share $\alpha_q$ declines sharply and becomes zero once the loading exceeds about $20\%$ (i.e., for $\varphi \ge 1.2)$, so that it is no longer optimal to purchase any insurance. 
At the same time, the durable-good share $\alpha_k$ decreases and then flattens out for higher loadings, and the reduction in durable holdings is absorbed by a higher allocation to the risky asset, so that $\alpha_\pi$ increases. 
By contrast, the consumption share $\alpha_c$ remains almost constant over the entire range of insurance loads. 

\begin{figure}[htbp] 
\centering 
\captionsetup{width=0.75\textwidth,format=plain} 
\includegraphics[width=0.75\linewidth]{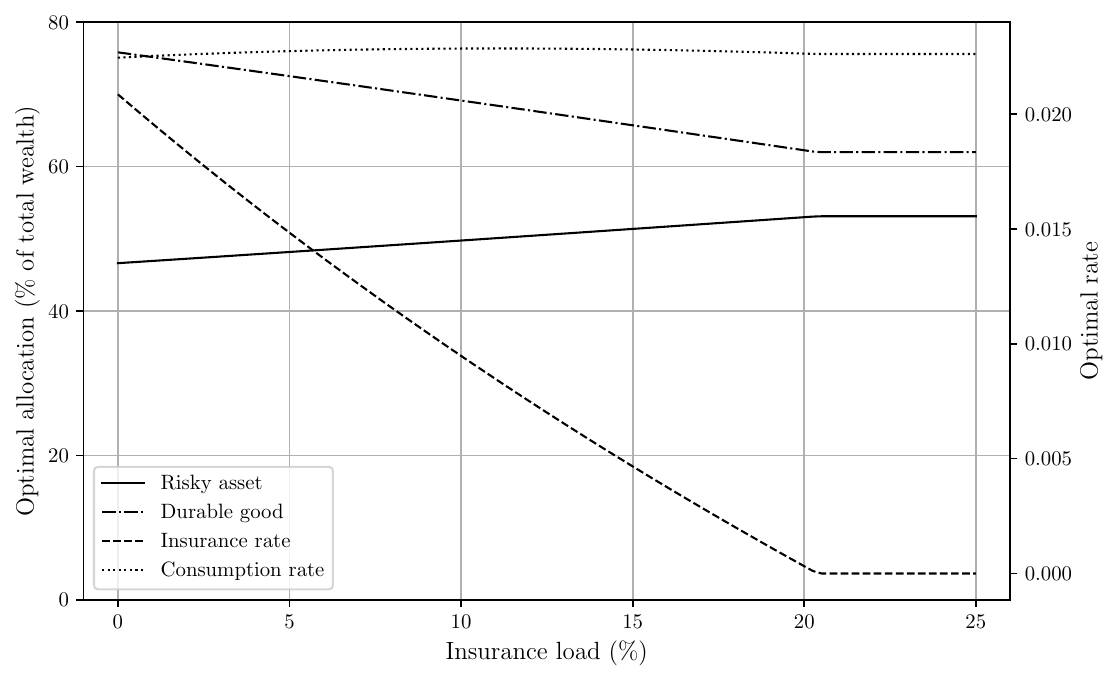} 
\caption{Optimal allocation for the base model under different insurance loads.} 
\label{sensitivity-loads} 
\end{figure} 

Figure~\ref{sensitivity-frequency} shows how the optimal allocations change as the annual damage frequency $\lambda_2$ varies. 
When damage events are rare, the agent holds a relatively large position in the durable good and finds it optimal either not to insure or to insure only modestly. 
As $\lambda_2$ increases, the optimal insurance share $\alpha_q$ rises, reflecting the higher marginal value of protection when insurable losses occur more frequently, while the allocation to the durable good $\alpha_k$ declines as repeated damage events make the durable good less attractive. 
After a certain point (for $\lambda_2 \approx 18\%$), the optimal insurance share $\alpha_q$ starts to decline, and becomes negligible once $\lambda_2$ reaches around 42$\%$. 
The consumption share $\alpha_c$ remains essentially flat over the range of $\lambda_2$ considered, while the share invested in the risky asset $\alpha_\pi$ is increasing, so that most of the adjustment takes place through the trade-off between durable holdings, insurance coverage, and the remaining wealth invested in financial assets. 

\begin{figure}[htbp] 
\centering 
\captionsetup{width=0.75\textwidth,format=plain} 
\includegraphics[width=0.75\linewidth]{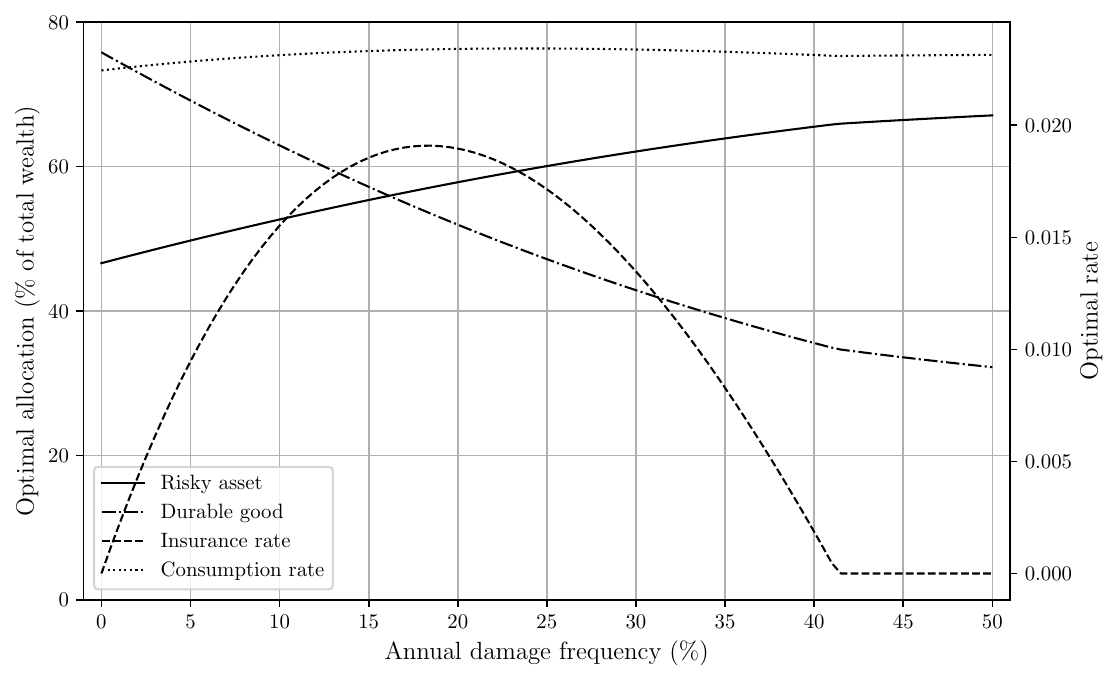} 
\caption{Optimal allocation for the base model under different values for the annual damage frequency $\lambda_2$.} 
\label{sensitivity-frequency} 
\end{figure} 

Figure~\ref{sensitivity-fraction} examines the sensitivity of the optimal strategy to the loss fraction $\ell$ incurred when an insurable event occurs. 
When the damage fraction is small ($\ell \le 0.25$), the optimal insurance coverage is zero, so the agent prefers to self-insure against minor losses rather than pay premiums. 
As $\ell$ increases beyond this threshold, the optimal insurance share $\alpha_k$ rises markedly, while the durable-good share is reduced in order to limit exposure to potentially large losses. 
In contrast, the risky-asset share $\alpha_\pi$ slightly increases with $\ell$, while the consumption share $\alpha_c$ remains essentially constant across the range of damage fractions considered. 

\begin{figure}[htbp] 
\centering 
\captionsetup{width=0.75\textwidth,format=plain} 
\includegraphics[width=0.75\linewidth]{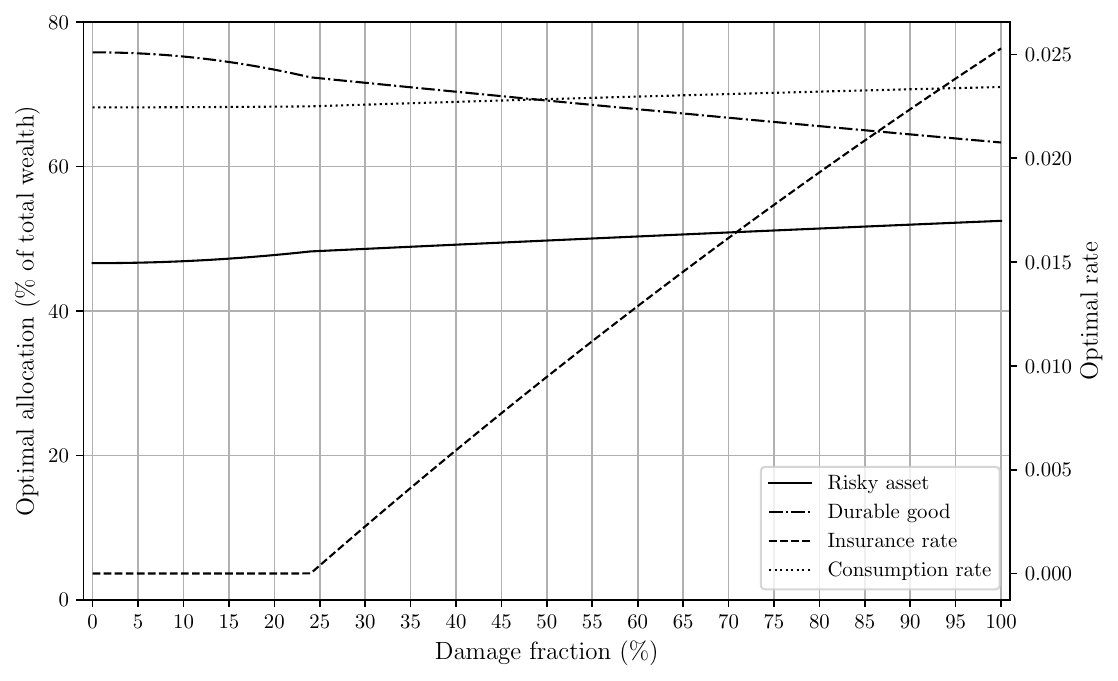} 
\caption{Optimal allocation for the base model under different values for the damage fraction $\ell$.} 
\label{sensitivity-fraction} 
\end{figure} 

Finally, Figure~\ref{sensitivity-aversion} shows the impact of risk aversion on the optimal allocation. 
As the risk aversion parameter $\gamma$ increases, the agent adopts a more conservative stance in financial markets: the share invested in the risky asset $\alpha_\pi$ declines, while the optimal insurance share $\alpha_q$ increases, so that a larger part of the durable-good risk is transferred to the insurer. 
The optimal insurance share $\alpha_q$ only increases up to a certain point (around $\gamma = 0.7$), after which it remains largely constant. 
The durable-good allocation $\alpha_k$ remains relatively high over the whole range of $\gamma$ and adjusts only moderately once the optimal insurance share $\alpha_q$ becomes effectively constant, indicating that the main margin of adjustment is the trade-off between risky investment and insurance, especially for moderate values of $\gamma$. 
The consumption rate, $\alpha_c$, while sharply increasing at the beginning, varies only mildly with $\gamma$ afterwards. 
Overall, the figure confirms that more risk-averse agents optimally hold less of the risky financial asset and demand more insurance coverage. 

\begin{figure}[htbp] 
\centering 
\captionsetup{width=0.75\textwidth,format=plain} 
\includegraphics[width=0.75\linewidth]{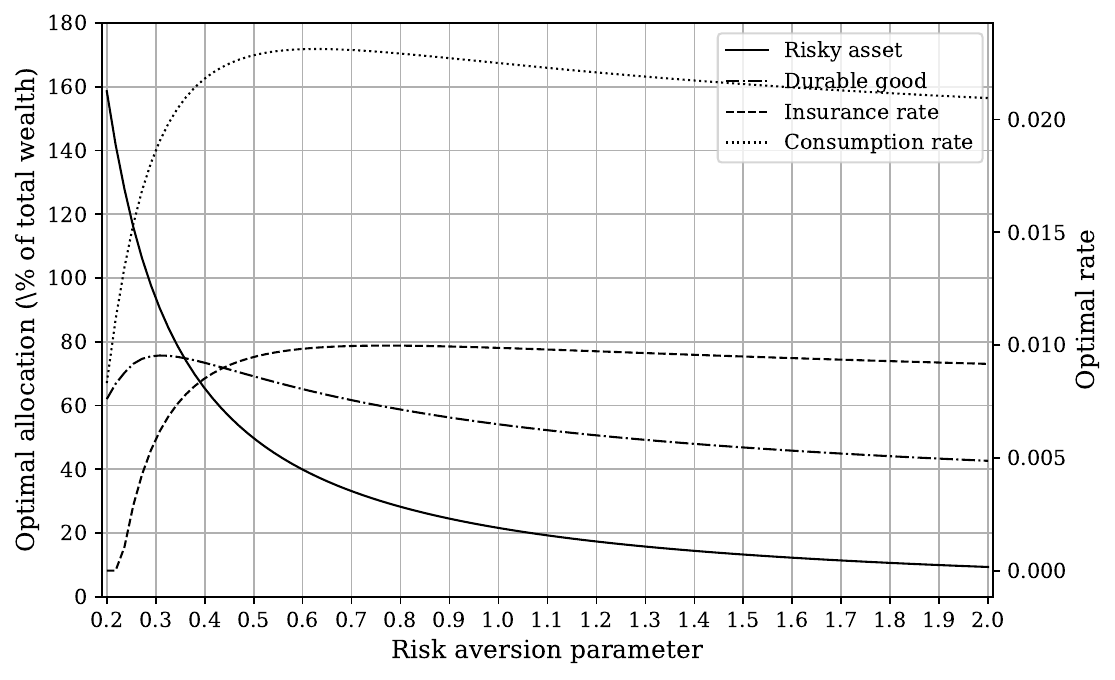} 
\caption{Optimal allocation for the base model under different values for the risk aversion factor $\gamma$.} 
\label{sensitivity-aversion} 
\end{figure} 

Table~\ref{tab:wealth_summary} shows that the insurance loading has a pronounced impact on the tails of the terminal-wealth distribution, even though the median remains essentially unchanged. 
Since all values are quoted in millions, the minimum wealth after one year drops from $0.34$ for $\varphi = 1.00$ to $0.23$ for $\varphi = 1.25$, a reduction of about $0.11$ million, i.e., more than $10\%$ of the initial wealth. 
Over the ten-year horizon, the right tail is also significantly affected: both the $99$th percentile and the maximum wealth decrease as $\varphi$ increases, with the maximum falling by more than one million when moving from actuarially fair pricing to $\varphi = 1.25$. 
These changes are mirrored in the higher moments: skewness and excess kurtosis decline markedly with the loading, both for $T=1$ and for $T=10$, indicating a substantial reduction in extreme positive outcomes and a visibly thinner, less skewed right tail when insurance becomes more expensive and is optimally used less aggressively. 

\begin{table}[htbp] 
\centering 
\begin{threeparttable} 
\caption{Distribution of terminal wealth $X(T)$ for different insurance loads $\varphi$ after one year ($T=1$) and after ten years ($T=10$), starting from $X(0) = 1$ million.} 
\label{tab:wealth_summary} 
\begin{tabular}{llrrrrrrrrr}
\toprule 
 $T$ & $\varphi$ & min & $q_1$ & $q_{25}$ & $q_{50}$ & $q_{75}$ & $q_{99}$ & max & skew & kurt \\ 
 \midrule 
 \multirow{3}{*}{$1$} & $1.00$ & 0.34 & 0.59 & 0.86 & 1.01 & 1.17 & 1.71 & 3.28 & 0.69 & 0.86 \\
 & $1.10$ & 0.33 & 0.59 & 0.87 & 1.01 & 1.17 & 1.68 & 3.16 & 0.65 & 0.76 \\
 & $1.25$ & 0.23 & 0.57 & 0.87 & 1.01 & 1.17 & 1.66 & 3.03 & 0.52 & 0.63 \\
 \midrule 
 \multirow{3}{*}{$10$} & $1.00$ & 0.04 & 0.20 & 0.65 & 1.06 & 1.73 & 5.70 & 30.32 & 3.01 & 18.49 \\
 & $1.10$ & 0.04 & 0.20 & 0.65 & 1.06 & 1.70 & 5.43 & 29.74 & 2.87 & 16.85 \\
 & $1.25$ & 0.03 & 0.19 & 0.65 & 1.06 & 1.71 & 5.39 & 29.06 & 2.75 & 15.36 \\
\bottomrule 
\end{tabular} 
\begin{tablenotes}[para] 
\footnotesize 
Note: Quantiles are quoted in millions. skew refers to skewness, and kurt refers to the excess kurtosis. 
\end{tablenotes} 
\end{threeparttable} 
\end{table}

\section{Conclusion}\label{conclusion_sec}
In this paper we considered an economic agent's optimal consumption and investment problem with an infinite time horizon in a continuous-time economy. 
The agent is faced with multiple risk factors including equity market risks and crashes, durable good price fluctuations and insurable events that cause damage to the durable good, and must make multiple decisions including risky and riskless asset allocations, perishable and durable goods consumption, and optional insurance coverage for the durable good. 
The agent derives utility from consumption over an infinite time horizon via a CRRA-type utility function. 

This paper is built upon \citet{damgaard2003} and \cite{matos2014}. 
The key innovation we consider is that the durable good is subject to random insurable losses modelled by a Poisson jump process. 
We proved concavity of the HJB function and used Karush--Kuhn--Tucker conditions for constrained optimisation to derive a semi-explicit solution for the optimal decision problem when no transaction costs are present. 
Only first-order conditions are considered in models studied in \cite{damgaard2003} and \cite{matos2014} without proof of concavity. 
Our model is a generalisation of these models and thus our proof also applies to these special cases.
We also proved a verification theorem that identifies sufficient conditions for the optimal strategy. 
Again, it can be applied to the special case considered in \cite{matos2014}, where a verification theorem was not proved. Finally, we considered risk aversion regimes $0<\gamma<1$ and empirically relevant $\gamma>1$, while \cite{matos2014} and \cite{damgaard2003} considered $0<\gamma<1$ only.

Our numerical results illustrate that the optimal allocations behave in an economically intuitive way and confirm the comparative statistics suggested by the closed-form solution. 
An increase in the premium loadings reduces both the demand for the durable good and the insurance coverage, while shifting wealth into the risky financial asset, with insurance demand vanishing altogether for sufficiently large loadings. 
Changes in the arrival intensity and severity of insurable events alter the trade-off between self-insurance and market insurance: moderate increases in frequency or loss fraction raise the optimal insurance coverage, whereas very high frequencies make the durable good itself less attractive. 
Varying risk aversion shows that more risk-averse agents optimally hold less of the risky asset and demand more insurance, while maintaining a substantial position in the durable good. 
Finally, the wealth distributions in Table~\ref{tab:wealth_summary} reveal that the insurance loading has a pronounced impact on the tails of long-run wealth, significantly reducing extreme positive outcomes and thinning the right tail when insurance is expensive. 

In this paper we did not incorporate transaction costs on durable goods, which are empirically important. 
Introducing insurable jump risk in the presence of proportional transaction costs substantially complicates the analysis, and a tractable treatment will likely require numerical methods. 
Extending the present framework along these lines appears to be an interesting direction for future research. 

\bibliographystyle{elsart-harv}
\bibliography{gmwxb}

\clearpage

\appendix
\section{Proof of verification theorem and related auxiliary results (Lemmas \ref{nobankruptcy}-\ref{Integrability} and Theorems \ref{verification}-\ref{verificationCand2})}

\noindent

\noindent
{\bf Proof of Lemma \ref{nobankruptcy}}

\begin{proof} 
Under the feedback controls \eqref{optimalcontrol} we have: 
\begin{equation*} 
c(t) = \alpha_c X(t), \quad \pi_1(t) = \alpha_\pi X(t), \quad K(t)P(t) = \alpha_k X(t), \quad q(t) = \alpha_q X(t). 
\end{equation*} 
Substituting these controls into the wealth dynamics \eqref{dxt} gives an SDE of the form:
\begin{equation*} 
dX(t) = X(t-)\Bigl(a\,dt + b_1\,dB_1(t) + b_2\,dB_2(t)\Bigr) + X(t-)\Bigl((\xi_1-1)\,dN_1(t) + (\xi_2-1)\,dN_2(t)\Bigr), 
\end{equation*} 
where $a, b_1, b_2$ are some real constants, and the jump amplitudes are given by:
\begin{equation*} 
\xi_1 = 1-\eta\alpha_\pi, \qquad \xi_2 = 1 - \ell\alpha_k + \alpha_q. 
\end{equation*} 
Define the semimartingale $\{ L (t) \}_{t \ge 0}$ by putting:
\begin{eqnarray} \label{L(t)}
d L (t) :=  a\,dt + b_1\,dB_1(t) + b_2\,dB_2(t) + (\xi_1-1)\,dN_1(t) + (\xi_2-1)\,dN_2(t)
\end{eqnarray}
Then 
\begin{eqnarray} \label{LinearSDE}
d X(t) = X (t-) d L (t).
\end{eqnarray}
Write $\mathcal{E}(L)$ for the Doléans--Dade stochastic exponential of the semimartingale $\{ L (t) \}_{t \ge 0}$. Then by \cite{jacod2003limit} (Chapter~II, Section~8a, Eq.~(8.2)), 
\begin{eqnarray} \label{StochasticExponential}
\mathcal{E}(L) (t) = \exp \left ( L (t) - \frac{1}{2} \left <  L^c, L^c \right > (t) \right ) \prod_{0 < u \le t} (1 + \Delta L (u)) \exp (- \Delta L (u) ),
\end{eqnarray}
where $L^c := \{ L^c (t) \}_{t \ge 0}$ is the continuous martingale part of $\{ L (t) \}_{t \ge 0}$; $\{ \left < L^c, L^c \right > (t) \}_{t \ge 0}$ is the quadratic variation of $L^c$. Then it is well known (see, e.g., \cite{jacod2003limit} (Chapter~I, Section~4f, Theorem~4.61)) that the unique solution of the linear SDE in Eq. (\ref{LinearSDE}) is given by:
\begin{eqnarray} \label{LinearSDESol}
X (t) = X(0) \mathcal{E}(L) (t), \quad \mbox{for each $t \ge 0$}, \quad \mbox{$\mathbb{P}$-a.s.}
\end{eqnarray}
Note that the solvency constraints \eqref{solvency1} imply $\xi_1>0$ and $\xi_2>0$. Consequently, the jump of $L (t)$ at time $t$ satisfies:
\begin{eqnarray} \label{LJump}
\Delta L (t) \ge \min\{\xi_1-1, \xi_2-1\} > - 1. 
\end{eqnarray}
By Eq. (\ref{StochasticExponential}), Eq. (\ref{LinearSDESol}) and Eq. (\ref{LJump}), $X(t)>0$, for all $t \ge 0$, $\mathbb{P}$-a.s.
This also implies that $X(t)^\theta$ is well defined and positive, for all $\theta\in\mathbb{R}$ and $t\ge0$, $\mathbb{P}$-a.s.
\end{proof} 

\noindent
{\bf Proof of Lemma \ref{Integrability}}

\begin{proof}
First, from (\ref{gvfunntc}), we have:
\begin{eqnarray} \label{V(R(t)}
| V (R(t)) | &=& | V(s+t, X(t), P(t)) | = \bigg | \frac{1}{1-\gamma} \ap_v e^{- \rho (s + t)} (P (t))^{A} (X (t))^{B} \bigg |  \nonumber\\
&=& e^{- \rho (s + t)} M_1 | (P (t))^{A} (X (t))^{B} |, 
\end{eqnarray}
where $M_1 := | \frac{1}{1 - \gamma} \alpha_v |$, which is a finite positive constant.

From (\ref{hjb0}), (\ref{optimalcontrol}) and (\ref{ARgenerator}), we have:
\begin{eqnarray} \label{HJBEq1}
{\cal A}^{(c, \pi_1, k, q)}_R [V(s, x, p)] + \frac{1}{1-\gamma} e^{- \rho s} ({\hat c}^{\beta} {\hat k}^{1 - \beta} )^{1-\gamma} = 0,
\end{eqnarray}
where ${\hat c} = \ap_c x$ and ${\hat k} = \ap_k (\frac{x}{p})$.

\noindent
Then for each $t \ge 0$,
\begin{eqnarray} \label{HJBEq2}
{\cal A}^{(c, \pi_1, k, q)}_R [V(s+t, X(t), P(t))] + \frac{1}{1-\gamma} e^{- \rho (s+t)} (c^{\beta} (t) k^{1 - \beta} (t) )^{1-\gamma} = 0,
\end{eqnarray}
where $c (t) = \ap_c X(t)$ and $k (t) = \ap_k (\frac{X(t)}{P(t)})$.

\noindent
Consequently, using (\ref{optc}), \eqref{optimalcontrol} and (\ref{HJBEq2}),
\begin{eqnarray} \label{AR}
| {\cal A}^{(c, \pi_1, k, q)}_R [V(R (t))] | &=& | {\cal A}^{(c, \pi_1, k, q)}_R [V(s+t, X(t), P(t))] |
= \bigg | \frac{1}{1-\gamma} e^{- \rho (s+t)} ({\hat c}^{\beta} (t) {\hat k}^{1 - \beta} (t) )^{1-\gamma} \bigg |  \nonumber\\
&=& \bigg | \frac{1}{1-\gamma} e^{- \rho (s+t)} 
\ap_c^{\beta (1 - \gamma)} \ap_k^{(1-\beta)(1-\gamma)} P(t)^{-(1-\beta)(1-\gamma)} X(t)^{1-\gamma} \bigg | \nonumber\\
&=& e^{- \rho (s+t)} M_2 | (P (t))^{A} (X (t))^{B} |, 
\end{eqnarray}
where $M_2 = |\frac{\ap_v \ap_c}{\beta (1 - \gamma)} |$, which is a finite positive constant. 

From Lemma \ref{nobankruptcy}, by setting $\theta = B$, $(X (t))^B \ge 0$, $\mathbb{P}$-a.s., $\forall t \ge 0$. 
Then, to prove that for any finite stopping time $\tau$, 
\begin{eqnarray} 
\mathbb{E}^{(s, x, p)} [ | V (R (\tau) | ] < \infty,
\end{eqnarray}
it suffices to prove that for any finite stopping time $\tau$,
\begin{eqnarray} \label{Expectation1} 
\mathbb{E}^{(s, x, p)} [  e^{- \rho \tau} (P (\tau))^{A} (X (\tau))^{B} ] < \infty,
\end{eqnarray}
Furthermore, to prove that for any finite stopping time $\tau$,
\begin{eqnarray} 
\mathbb{E}^{(s, x, p)} \left [ \int^{\tau}_{0} |
{\cal A}^{(c, \pi_1, k, q)}_R [V (R (t)) ] | d t \right ] < \infty,
\end{eqnarray}
it suffices to prove that 
for any finite stopping time $\tau$,
\begin{eqnarray} \label{Expectation2}
\mathbb{E}^{(s, x, p)} \bigg [ \int^{\tau}_0 e^{- \rho t} (P (t))^{A} (X(t))^{B} d t \bigg ] < \infty.
\end{eqnarray}
We shall first prove (\ref{Expectation1}). To this end, let $Y (t) = (P (t))^{A} (X (t))^{B}$, for each $t \ge 0$.

\noindent
Applying It\^o's formula to $(P (t))^{A}$ gives:
\beqr{PA}
(P (t))^{A}  =& (P (0))^{A} + \int^{t}_{0} A (P (u))^{A} \bigg (\mu_P + \frac{1}{2} (A-1) (\sigma^2_1 + \sigma^2_2) \bigg ) d u \\
&+ \int^{t}_{0} A (P (u))^{A} \sigma_1 d B_1 (u) + \int^{t}_{0} A (P (u))^{A} \sigma_2 d B_2 (u). 
\eeqr
Applying It\^o's formula for semimartingales to $(X (t))^{B}$ and using (\ref{optimalcontrol}) give:
\sloppy
\beqr{XB}
(X (t))^{B} =& (X (0))^{B} + \int^{t}_{0} (X (u))^{B} \bigg ( B (r + \ap_{\pi} (\mu_S + \lambda_1 \eta - r) + \ap_k
(\mu_P - r - \delta) \\
& - \varphi \lambda_2 \ap_q - \ap_c) + \frac{1}{2} B(B - 1) ((\ap_{\pi} \sigma_S + \ap_k \sigma_1)^2 + \ap^2_k \sigma^2_2) + ((1 - \ap_{\pi} \eta)^{B} - 1) \lambda_1 \\
& + ((1 - \ap_k l + \ap_q)^{B} - 1) \lambda_2 \bigg ) du  
+ \int^{t}_{0} B (X (u))^{B} (\ap_{\pi} \sigma_S + \ap_k \sigma_1) d B_1 (u) \\
& + \int^{t}_{0} B (X (u))^{B} \ap_k \sigma_2 d B_2 (u) + 
\int^{t}_{0} (X (u-))^{B} ( (1 - \ap_{\pi} \eta)^{B} - 1)
d {\tilde N}_1 (u) \\
& + \int^{t}_{0} (X (u-))^{B} ((1 - \ap_k l + \ap_q)^{B} - 1) 
d {\tilde N}_2 (u),
\eeqr
where $\{ {\tilde N}_1 (t) \}_{t \ge 0}$ and
$\{ {\tilde N}_2 (t) \}_{t \ge 0}$ are two compensated
Poisson processes such that ${\tilde N}_1 (t) := N_1 (t) - \lambda_1 t$ and ${\tilde N}_2 (t) := N_2 (t) - \lambda_2 t$.

\noindent
By the It\^o's product rule, 
\beqr{PAXB}
(P (t))^{A} (X (t))^{B} =& (P (0))^{A} (X (0))^{B}
+ \int^{t}_{0} (P(u-))^{A} d (X (u))^{B} + 
\int^{t}_{0} (X(u-))^{B} d (P (u))^{A} \\
& + [P^{A}, X^{B}] (t),
\eeqr 
where $\{ [P^{A}, X^{B}] (t) \}_{t \ge 0}$ is the 
(optional) quadratic co-variation between $\{ (P (t))^{A} \}_{t \ge 0}$ and $\{ (X (t))^{B} \}_{t \ge 0}$.

Write, for each $t \ge 0$, $\Delta P^{A} (t) = (P (t))^{A} - (P (t-))^{A}$, (i.e., the jump of $(P (t))^{A}$ at time $t$),
and $\Delta X^{B} (t) = (X (t))^{B} - (X (t-))^{B}$, (i.e., the jump of $(X (t))^{B}$ at time $t$). For simplicity, it is 
supposed that $\Delta P^{A} (0) = \Delta X^{B} (0) = 0$,
(i.e., there are no jumps at time $0$ for the two processes
$\{ (P (t))^{A} \}_{t \ge 0}$ and $\{ (P (t))^{B} \}_{t \ge 0}$).
Since $\{ (P (t))^{A} \}_{t \ge 0}$ is a continuous process,
$\Delta P^{A} (t) = 0$, $\forall t \ge 0$. Consequently,
\begin{eqnarray} \label{OQV}
[ P^{A}, X^{B} ] (t) &=& \left < (P^{A})^c, (X^{B})^c \right > (t) 
+ \sum_{0 < u \le t} \Delta P^{A} (u) \Delta X^{B} (u) \nonumber \\
&=& \left < (P^{A})^c, (X^{B})^c \right > (t), 
\end{eqnarray}
where $\{ (P^{A})^c (t) \}_{t \ge 0}$ is the continuous
martingale part of $\{ (P (t))^{A} \}_{t \ge 0}$
and $\{ (X^{B})^c (t) \}_{t \ge 0}$ is the continuous
martingale part of $\{ (X (t))^{B} \}_{t \ge 0}$;
$\{ \left < (P^{A})^c, (X^{B})^c \right > (t) \}_{t \ge 0}$
is the predictable quadratic co-variation between 
$\{ (P^{A})^c (t) \}_{t \ge 0}$ and $\{ (X^{B})^c (t) \}_{t \ge 0}$. 

\noindent
Then 
\begin{eqnarray} \label{PQV}
\left < (P^{A})^c, (X^{B})^c \right > (t) = \int^{t}_{0} 
AB (P (u))^{A} (X (u))^{B} (\ap_{\pi} \sigma_S \sigma_1 + \ap_k (\sigma^2_1 + \sigma^2_2) ) d u.
\end{eqnarray}
Consequently, using (\ref{T1}), (\ref{PA}), (\ref{XB}), (\ref{PAXB}), (\ref{OQV}) and (\ref{PQV}), 
\begin{eqnarray} \label{PAXB1}
(P (t))^{A} (X (t))^{B} &=& (P (0))^{A} (X (0))^{B}
+ \int^{t}_{0} (P (u))^{A} (X (u))^{B} T_1 d u \nonumber \\
&& + \int^{t}_{0} (P (u))^{A} (X (u))^{B} (B \ap_{\pi} \sigma_S 
+ (B \ap_k + A) \sigma_1) d B_1 (u) \nonumber \\
&& + \int^{t}_{0} (P (u))^{A} (X (u))^{B} ((B \ap_k + A) \sigma_2) 
d B_2 (u) \nonumber \\
&& + 
\int^{t}_{0} (P (u-))^{A} (X (u-))^{B} ( (1 - \ap_{\pi} \eta)^{B} - 1)
d {\tilde N}_1 (u) \nonumber \\
&& + \int^{t}_{0} (P (u-))^{A} (X (u-))^{B} ((1 - \ap_k l + \ap_q)^{B} - 1) 
d {\tilde N}_2 (u) \nonumber\\
 &=& (P (0))^{A} (X (0))^{B}
+ \int^{t}_{0} (P (u))^{A} (X (u))^{B} T_1 d u \nonumber \\
&& + \int^{t}_{0} (P (u))^{A} (X (u))^{B} T_2 d B_1 (u) \nonumber \\
&& + \int^{t}_{0} (P (u))^{A} (X (u))^{B} T_3 d B_2 (u) \nonumber \\
&& + 
\int^{t}_{0} (P (u-))^{A} (X (u-))^{B} T_4
d {\tilde N}_1 (u) \nonumber \\
&& + \int^{t}_{0}  (P (u-))^{A} (X (u-))^{B} T_5
d {\tilde N}_2 (u), \ \mbox{say},
\end{eqnarray}
where
\begin{eqnarray*}
T_2 &:=& B \alpha_{\pi} \sigma_S + (B \alpha_k + A) \sigma_1, \nonumber\\
T_3 &:=& (B \alpha_k + A) \sigma_2, \nonumber\\
T_4 &:=& (1 - \alpha_{\pi} \eta)^{B} - 1, \nonumber\\
T_5 &:=& (1 - \alpha_k l + \alpha_q)^{B} - 1.
\end{eqnarray*}

Recall that $Y(t) = (P(t))^{A} (X(t))^{B}$, for $t \ge 0$. Then, from (\ref{PAXB1}), 
\begin{eqnarray} \label{Y}
Y(t) &=& Y(0) + \int^{t}_{0} Y(u) T_1 d u + \int^{t}_{0} Y(u) T_2 d B_1 (u) + \int^{t}_{0} Y(u) T_3 d B_2 (u) \nonumber \\
&& + \int^{t}_{0} Y(u-) T_4 d {\tilde N}_1 (u) + \int^{t}_{0} Y(u-) T_5 d {\tilde N}_2 (u).
\end{eqnarray}
Define another process $\{ Z (t) \}_{t \ge 0}$ such 
that $Y (t) = Y(0) \exp (Z(t))$.
Then applying It\^o's formula for semimartingales
to $\ln (Y(t)/Y(0))$ gives:
\begin{eqnarray} \label{Z}
Z (t) &=& \int^{t}_{0} \frac{1}{Y(u)} d Y (u) - \frac{1}{2} \int^{t}_{0} \frac{1}{Y^2 (u)} d \left < Y^c, Y^c \right > (u) \nonumber\\
&& + \sum_{0 < u \le t} \bigg ( \ln (Y(u-) + \Delta Y (u)) - \ln (Y(u-)) - \frac{1}{Y(u)} \Delta Y(u) \bigg ),
\end{eqnarray}
where $\Delta Y (t) = Y(t) - Y(t-)$; $\{ Y^c (t) \}_{t \ge 0}$ is the continuous martingale part of $\{ Y (t) \}_{t \ge 0}$. \\

\noindent
Recall that ${\tilde N}_1 (t) = N_1 (t) - \lambda_1 t$
and ${\tilde N}_2 (t) = N_2 (t) - \lambda_2 t$. Then
from (\ref{Y}), 
\begin{eqnarray} \label{Zc1}
&& \int^{t}_{0} \frac{1}{Y(u)} d Y(u) - \sum_{0 < u \le t} \frac{1}{Y(u)} \Delta Y(u) \nonumber\\
&=& \int^{t}_{0} (T_1 - T_4 \lambda_1
- T_5 \lambda_2) d u + \int^{t}_{0} T_2 d B_1 (u) +
\int^{t}_{0} T_3 d B_2 (u). 
\end{eqnarray}
Also,
\begin{eqnarray} \label{Zc2}
d \left < Y^c, Y^c \right > (t) = Y^2 (t) (T^2_1 + T^2_2) d t.
\end{eqnarray}
Since $\{ N_1 (t) \}_{t \ge 0}$ and $\{ N_2 (t) \}_{t \ge 0}$
are independent, $\Delta N_1 (t) \Delta N_2 (t) = 0$, $\forall t \ge 0$, (i.e., there are no common jumps between $\{ N_1 (t) \}_{t \ge 0}$ and $\{ N_2 (t) \}_{t \ge 0}$). Consequently,
\begin{eqnarray} \label{Zc3}
\ln (Y(t-) + \Delta Y(t)) - \ln (Y(t-)) = \ln (1 + T_4) \Delta N_1 (t) + \ln (1 + T_5) \Delta N_2 (t).
\end{eqnarray}
From (\ref{Z}), (\ref{Zc1}), (\ref{Zc2}), (\ref{Zc3}),
\begin{eqnarray} \label{Z1}
Z (t) &=& \int^{t}_{0} \bigg ( T_1 - \frac{1}{2} (T^2_2 + T^2_3)
+ \lambda_1 (\ln(1 + T_4) - T_4) + \lambda_2 (\ln(1 + T_5) - T_5) \bigg ) d u \nonumber\\
&& + \int^{t}_{0} T_2 d B_1 (u) + \int^{t}_{0} T_3 d B_2 (u) + \int^{t}_{0} \ln (1 + T_4) d {\tilde N}_1 (u) \nonumber\\
&& + \int^{t}_{0} \ln (1 + T_5) d {\tilde N}_2 (u) \nonumber\\
&=& \bigg ( T_1 - \frac{1}{2} (T^2_2 + T^2_3)
+ \lambda_1 (\ln(1 + T_4) - T_4) + \lambda_2 (\ln(1 + T_5) - T_5) \bigg ) t \nonumber\\
&& + T_2 B_1 (t) + T_3 B_2 (t) + \ln (1 + T_4) {\tilde N}_1 (t) + \ln (1 + T_5) {\tilde N}_2 (t).
\end{eqnarray}
Since $\{ B_1 (t) \}_{t \ge 0}$, $\{ B_2 (t) \}_{t \ge 0}$, $\{ N_1 (t) \}_{t \ge 0}$ and $\{ N_2 (t) \}_{t \ge 0}$ are
independent, for any finite stopping time $\tau$,
\begin{eqnarray} \label{mgfZ}
&& \mathbb{E}^{(s, x, p)} [ e^{- \rho \tau} Y (\tau)] \nonumber\\
&=& Y(0) \mathbb{E}^{(s, x, p)} [ e^{- \rho \tau} \exp (Z (\tau))] \nonumber\\
&=& Y(0) \mathbb{E}^{(s, x, p)} \bigg [ \exp \bigg ( (- \rho + T_1 + \lambda_1 (\ln(1 + T_4) - T_4) + \lambda_2 (\ln(1 + T_5) - T_5) ) \tau \bigg ) \bigg ]
\nonumber\\
&& 
\times \mathbb{E}^{(s, x, p)} \bigg [ \exp \bigg ( \ln (1 + T_4) {\tilde N}_1 (\tau) \bigg ) \bigg ] \mathbb{E}^{(s, x, p)} \bigg [ \exp \bigg ( \ln (1 + T_5) {\tilde N}_2 (\tau) \bigg ) \bigg ]. \nonumber\\
\end{eqnarray}
As in \cite{oksendal2019applied},
Exercise 1.6 therein, it can be shown that for any finite stopping time,
\begin{eqnarray} \label{mgfN1}
\mathbb{E}^{(s, x, p)} \bigg [ \exp \bigg ( \ln (1 + T_4) {\tilde N}_1 (\tau) \bigg ) \bigg ] = \mathbb{E}^{(s, x, p)} \bigg [ \exp \bigg ( (T_4 - \ln (1 + T_4) ) \lambda_1 \tau \bigg ) \bigg ] < \infty,
\end{eqnarray}
and that
\begin{eqnarray} \label{mgfN2}
\mathbb{E}^{(s, x, p)} \bigg [ \exp \bigg ( \ln (1 + T_5) {\tilde N}_2 (\tau) \bigg ) \bigg ] = \mathbb{E}^{(s, x, p)} \bigg [ \exp \bigg ( (T_5 - \ln (1 + T_5) ) \lambda_2 \tau \bigg ) \bigg ] < \infty. 
\end{eqnarray}
Consequently, using (\ref{mgfZ}), for any finite stopping time $\tau$, 
(\ref{mgfN1}), (\ref{mgfN2}), 
\begin{eqnarray} \label{mgfZ1Eq}
\mathbb{E}^{(s, x, p)} [ e^{- \rho \tau} Y (\tau)] &=& Y(0) \mathbb{E}^{(s, x, p)} \bigg [ \exp \bigg ( (- \rho + T_1 ) \tau \bigg ) \bigg ].
\end{eqnarray}
Since $\rho > T_1$,
\begin{eqnarray} \label{mgfZ1}
\mathbb{E}^{(s, x, p)} [ e^{- \rho \tau} Y (\tau)] \le K^{\dagger},
\end{eqnarray}
for some finite constant $K^{\dagger} > 0$. 
Then (\ref{Expectation1}) follows from (\ref{mgfZ1}).

It remains to prove (\ref{Expectation2}).
From (\ref{mgfZ1Eq}),
\begin{eqnarray} \label{EYtilde}
\mathbb{E}^{(s, x, p)} [ Y (t) ] = Y(0) \mathbb{E}^{(s, x, p)} [ \exp (T_1 t ) ] = Y(0) \exp (T_1 t ). 
\end{eqnarray}
Recall the condition that $\rho > T_1$.
Then using Fubini's theorem and (\ref{EYtilde}), we have:
\begin{eqnarray} \label{EPAXB1}
\mathbb{E}^{(s, x, p)} \bigg [ \int^{\tau}_0 e^{- \rho t} 
(P (t))^{A} (X (t))^{B}  d t \bigg ] &\le&
\mathbb{E}^{(s, x, p)} \bigg [ \int^{\infty}_0 e^{- \rho t} 
(P (t))^{A} (X (t))^{B}  d t \bigg ] \nonumber\\
&=& \int^{\infty}_0 e^{- \rho t} \mathbb{E}^{(s, x, p)} [Y (t)] d t \nonumber\\
&=& Y(0) \int^{\infty}_0 \exp \bigg ((T_1 - \rho) t \bigg ) d t \nonumber\\
&=& \frac{Y(0)} {T_1 - \rho} \lim_{t \to \infty} \bigg ( e^{(T_1 - \rho) t} - 1 \bigg ) \nonumber\\
&=& \frac{Y(0)} {\rho - T_1} < \infty,
\end{eqnarray}
This completes the proof of the lemma. 
\end{proof}

\noindent
{\bf Proof of Theorem \ref{verification}}

\begin{proof} The proof is similar to that of Theorem 1 in Framstad et al. (2001b). First, note that $T_L \coloneqq \min \{ L, \inf \{ t > 0 \colon |X(t) | \ge L \} \} \le L < \infty$. Then using the Dynkin formula and Conditions (a)(i) and (a)(ii), 
\begin{eqnarray} \label{Dynkin1}
 \mathbb{E}^{(s, x, p)} \left [ e^{- \rho T_L} \nu (R (T_L)) \right ] 
&=& 
\nu (s, x, p) + \mathbb{E}^{(s, x, p)} \left [ \int^{T_L}_{0} e^{- \rho t} {\cal A}^{(c, \pi_1, k, q)} [\nu (R (t))] d t \right ] \nonumber\\
&\le& \nu (s, x, p) - \mathbb{E}^{(s, x, p)} \left [ \int^{T_L}_{0} e^{- \rho t} u (c(t), k(t)) d t \right ].
\end{eqnarray}
This implies that 
\begin{eqnarray} \label{Dynkin2}
\nu (s, x, p) \ge \mathbb{E}^{(s, x, p)} \left [ \int^{T_L}_{0} e^{- \rho t} u (c(t), k(t)) d t + e^{- \rho T_L} \nu (R (T_L)) \right ].
\end{eqnarray}
Consequently,
\begin{eqnarray} \label{Dynkinliminf}
\nu (s, x, p) \ge \liminf_{L \to \infty} \left \{ \mathbb{E}^{(s, x, p)} \left [ \int^{T_L}_{0} e^{- \rho t} u (c(t), k(t)) d t + e^{- \rho T_L} \nu (R (T_L)) \right ] \right \}.
\end{eqnarray}
Given that $\nu$ is a non-negative function. Then by the Fatou lemma 
and Eq. (\ref{Dynkinliminf}),
\begin{eqnarray} \label{Fatou}
\nu (s, x, p) &\ge& \liminf_{L \to \infty} \left \{ \mathbb{E}^{(s, x, p)} \left [ \int^{T_L}_{0} e^{- \rho t} u (c(t), k(t)) d t + e^{- \rho T_L} \nu (R (T_L)) \right ] \right \} \nonumber\\
&\ge&  \mathbb{E}^{(s, x, p)} \left [ \liminf_{L \to \infty} \left (\int^{T_L}_{0} e^{- \rho t} u (c(t), k(t)) d t + e^{- \rho T_L} \nu (R (T_L)) \right ) \right ] \nonumber\\
&=& \mathbb{E}^{(s, x, p)} \left [ \int^{\infty}_{0} e^{- \rho t} u (c(t), k(t)) d t \right ] \nonumber\\
&=& J^{(c, \pi_1, k, q)} (s, x, p), \quad \forall (c, \pi_1, k, q).
\end{eqnarray}
Consequently, 
\begin{eqnarray} \label{ProofParta}
\nu (s, x, p) \ge \sup_{(c, \pi_1, k, q)} J^{(c, \pi_1, k, q)} (s, x, p) = V(s, x, p).
\end{eqnarray}
This proves the conclusion in Part (a).

It remains to prove the conclusion in Part (b). Again using the Dynkin formula and Condition (b)(i),
\begin{eqnarray} \label{DynkinPartb}
\mathbb{E}^{(s, x, p)} \left [ e^{- \rho T_L} \nu (R (T_L)) \right ] 
&=& 
\nu (s, x, p) + \mathbb{E}^{(s, x, p)} \left [ \int^{T_L}_{0} e^{- \rho t} {\cal A}^{({\hat c}, {\hat \pi}_1, {\hat k}, {\hat q})} [\nu (R (t))] d t \right ] \nonumber\\
&=& \nu (s, x, p) - \mathbb{E}^{(s, x, p)} \left [ \int^{T_L}_{0} e^{- \rho t} u ({\hat c}(t), {\hat k}(t)) d t \right ].
\end{eqnarray}
This implies that
\begin{eqnarray} \label{DynkinPartb1}
\nu (s, x, p) = \mathbb{E}^{(s, x, p)} \left [ \int^{T_L}_{0} e^{- \rho t} u ({\hat c}(t), {\hat k}(t)) d t + e^{- \rho T_L} \nu (R (T_L)) \right ]. 
\end{eqnarray}
By Condition (b)(iii) and Eq. (\ref{DynkinPartb1}),
\begin{eqnarray} \label{DynkinPartb1limit}
\nu (s, x, p) &=& \lim_{L \to \infty} \mathbb{E}^{(s, x, p)} \left [ \int^{T_L}_{0} e^{- \rho t} u ({\hat c}(t), {\hat k}(t)) d t \right ] \nonumber\\
&\le& \mathbb{E}^{(s, x, p)} \left [ \int^{\infty}_{0} e^{- \rho t} u ({\hat c}(t), {\hat k}(t)) d t \right ].
\end{eqnarray}
The last inequality follows from the fact that the utility function $u$ is non-negative when $\gamma \in (0, 1)$. Therefore, from Eq. (\ref{DynkinPartb1limit}),
\begin{eqnarray} \label{ProofPartb}
\nu (s, x, p) 
\le \mathbb{E}^{(s, x, p)} \left [ \int^{\infty}_{0} e^{- \rho t} u ({\hat c}(t), {\hat k}(t)) d t \right ] = V(s, x, p).
\end{eqnarray}
This, together with the conclusion in Part (a), proves the conclusion in Part (b).
\end{proof}

\noindent
{\bf Proof of Theorem \ref{verificationCand1}}

\noindent
Note that $1 - \gamma > 0$ and that ${\hat \alpha}_v \ge 0$. For $(s, x, p) \in \Re_{\ge 0} \times {\cal O}$, the candidate value function $V (s, x, p) \ge 0$ when $\gamma < 1$.
It is obvious that the candidate value function $V (s, x, p) \in {\cal C}^{1, 2} (\Re_{\ge 0} \times {\cal O})$. By Eq. (\ref{hjba}) and Eq. (\ref{vfunntc}), the candidate value function $V (s, x, p)$ satisfies:
\begin{eqnarray}
{\cal A}^{(c, \pi_1, k, q)} [V (s, x, p)] + u (c, k) \le 0, \quad \forall (c, \pi_1, k, q) \in \Re_{> 0} \times \Re \times \Re_{> 0} \times \Re_{\ge 0}.
\end{eqnarray}
That is, the candidate value function $V (s, x, p)$ satisfies
Condition (a)(i) in Theorem \ref{verification}. By Lemma \ref{Integrability}, the candidate value function $V (s, x, p)$ satisfies
Condition (a)(ii) in Theorem \ref{verification}.

It remains to prove that the candidate value function $V (s, x, p)$ satisfies Conditions (b)(i), (b)(ii) and (b)(iii) in Theorem \ref{verification}. By Eq. (\ref{hjba}), Eq. (\ref{optc}), Eq. (\ref{optc0}), Eq. (\ref{vfunntc}), and the fact that $H (\alpha_c, \alpha_{\pi}, \alpha_k, \alpha_q)$ is a concave function,
Condition (b)(i) in Theorem \ref{verification} is satisfied. It is obvious that Condition (b)(ii) in Theorem \ref{verification} is satisfied since $({\hat \alpha}_c, {\hat \alpha}_{\pi_1}, {\hat \alpha}_k, {\hat \alpha}_q)$ satisfies the constraints in Eq. (\ref{param_constraints}).

\noindent
It remains to check the transversality condition (Condition (b)(iii)) of Theorem \ref{verification}. To prove that Condition (b)(iii) holds, we recall that $Y (t) = (P (t))^A (X (t))^B$, for each $t \ge 0$. Then from the proof of Lemma \ref{Integrability}, it can be shown that
\begin{eqnarray} \label{mgfZVer}
&& \mathbb{E}^{(s, x, p)} [ e^{- \rho T_L} Y (T_L)] \nonumber\\
&=& Y(0) \mathbb{E}^{(s, x, p)} \bigg [ \exp \bigg ( (- \rho + T_1 + \lambda_1 (\ln(1 + T_4) - T_4) + \lambda_2 (\ln(1 + T_5) - T_5) ) T_L \bigg ) \bigg ]
\nonumber\\
&& 
\times \mathbb{E}^{(s, x, p)} \bigg [ \exp \bigg ( \ln (1 + T_4) {\tilde N}_1 (T_L) \bigg ) \bigg ] \mathbb{E}^{(s, x, p)} \bigg [ \exp \bigg ( \ln (1 + T_5) {\tilde N}_2 (T_L) \bigg ) \bigg ]. \nonumber\\
\end{eqnarray}
Note that $T_L \coloneqq \min (L, \inf \{ t > 0 \colon | {\hat Z} (t) | \le L \}) \le L < \infty$. As in \cite{oksendal2019applied},
Exercise 1.6 therein, it can be shown that for $T_L < \infty$,
\begin{eqnarray} \label{mgfN1Ver}
\mathbb{E}^{(s, x, p)} \bigg [ \exp \bigg ( \ln (1 + T_4) {\tilde N}_1 (T_L) \bigg ) \bigg ] = \mathbb{E}^{(s, x, p)} \bigg [ \exp \bigg ( (T_4 - \ln (1 + T_4) ) \lambda_1 T_L \bigg ) \bigg ] < \infty,
\end{eqnarray}
and that
\begin{eqnarray} \label{mgfN2Ver}
\mathbb{E}^{(s, x, p)} \bigg [ \exp \bigg ( \ln (1 + T_5) {\tilde N}_2 (T_L) \bigg ) \bigg ] = \mathbb{E}^{(s, x, p)} \bigg [ \exp \bigg ( (T_5 - \ln (1 + T_5) ) \lambda_2 T_L \bigg ) \bigg ] < \infty. 
\end{eqnarray}
Consequently, 
\begin{eqnarray} \label{mgfZ1Eq}
\mathbb{E}^{(s, x, p)} [ e^{- \rho T_L} Y (T_L)] &=& Y(0) \mathbb{E}^{(s, x, p)} \bigg [ \exp \bigg ( (- \rho + T_1 ) T_L \bigg ) \bigg ].
\end{eqnarray}
Since $T_1 < \rho$, 
\begin{eqnarray} \label{TC}
\lim_{L \to \infty} \mathbb{E}^{(s, x, p)} [e^{- \rho T_L} Y (T_L)] = Y (0) \lim_{L \to \infty} \mathbb{E}^{(s, x, p)} \bigg [ \exp \bigg ( (- \rho + T_1 ) T_L \bigg ) \bigg ] = 0.
\end{eqnarray}
Recall that
\begin{eqnarray} \label{ValueFunctionVer}
\nu (R (T_L)) = \frac{1}{1 - \gamma} \alpha_v e^{- \rho s} e^{- \rho T_L} Y (T_L).
\end{eqnarray}
Then 
\begin{eqnarray} \label{ValueFunctionVer1}
0 \le \mathbb{E}^{(s, x, p)} [ e^{-\rho T_L} \nu (R (T_L)) ] \le \frac{1}{1 - \gamma} \alpha_v  \mathbb{E}^{(s, x, p)} [e^{- \rho T_L} Y (T_L)].
\end{eqnarray}
By sending $L \to \infty$ in Eq. (\ref{ValueFunctionVer1}) and using Eq. (\ref{TC}), we have:
\begin{eqnarray} \label{ValueFunctionVer2}
0 \le \lim_{L \to \infty} \mathbb{E}^{(s, x, p)} [ e^{-\rho T_L} \nu (R (T_L)) ] \le \frac{1}{1 - \gamma} \alpha_v \lim_{L \to \infty} \mathbb{E}^{(s, x, p)} [e^{- \rho T_L} Y (T_L)] = 0.
\end{eqnarray}
Consequently,
\begin{eqnarray} \label{ValueFunctionVer3}
\lim_{L \to \infty} \mathbb{E}^{(s, x, p)} [ e^{-\rho T_L} \nu (R (T_L)) ] = 0.
\end{eqnarray}
This proves that Condition (b)(iii) in Theorem \ref{verification} holds.

\noindent
{\bf Proof of Theorem \ref{verificationCand2}}

\begin{proof} We need to check that Conditions (i)-(vi) 
in Theorem 5.1 of \cite{oksendal2019applied} are satisfied
by the candidate value function in (\ref{gvfunntc}) and the candidate optimal control processes in (\ref{optimalcontrol}). 

\noindent
First, it is obvious that the candidate value function in (\ref{gvfunntc})  $V (s, x, p) \in {\cal C}^2 (\Re_{\ge 0} \times {\cal O}) \cap {\cal C} (\overline{\Re_{\ge 0} \times {\cal O}})$, and 
the candidate optimal control processes in  (\ref{optimalcontrol}) are in ${\cal A}$. Using 
the HJB equation in (\ref{hjb0}), the candidate optimal control processes in (\ref{optimalcontrol}), the optimisation 
problem in (\ref{hjba}) and the concavity of the function
$H (\ap_c, \alpha_{\pi},\ap_k,\ap_{q})$ to be proved
in \ref{ConcavityProof_sec}, Conditions (i) and (v)
in Theorem 5.1 of \cite{oksendal2019applied}
are satisfied.

Let $\tau_{\cal O} = \inf \{ t \ge 0\, |\, (X(t), P(t)) \notin {\cal O} \}$. To prove that Condition (ii) in Theorem
5.1 of \cite{oksendal2019applied}
is satisfied, it suffices
to prove that
\begin{eqnarray} \label{limit}
\lim_{t \to \tau_{\cal O}-} V (R(t)) = 0, \quad \mbox{$\mathbb{P}$-a.s.}
\end{eqnarray} 
Note that from the definition of $\tau_{\cal O}$, for all $t \in [0, \tau_{\cal O}-)$, $X (t) \in (0, \infty)$ and $P (t) \in (0, \infty)$. Then, for all $t \in [0, \tau_{\cal O}-)$, $(X (t))^B \in (0, \infty)$ and $(P (t))^A \in (0, \infty)$.
Furthermore, from Eq. (\ref{PositiveX(t)}) in Lemma \ref{nobankruptcy}, $\tau_{\cal O} = \infty$, $\mathbb{P}$-a.s. Consequently,
\begin{eqnarray} \label{limit2}
\lim_{t \to \tau_{\cal O}-} V (R(t)) &=& \lim_{t \to \tau_{\cal O}-} V (s+t, X(t), P(t)) \nonumber\\
&=& \lim_{t \to \tau_{\cal O}-} \bigg ( \frac{1}{1-\gamma} \alpha_v e^{-\rho (s+t)} (P (t))^A (X (t))^B \bigg ) = 0, \quad \mbox{$\mathbb{P}$-.a.s.}
\end{eqnarray}
This is because $\lim_{t \to \tau_{\cal O}-} e^{-\rho (s+t)} = \lim_{t \to \infty} e^{-\rho (s+t)} = 0$, $\mathbb{P}$-a.s.
Consequently, Condition (ii) in Theorem
5.1 of \cite{oksendal2019applied} is proved. 

Under Condition (ii) of this theorem, (say $\rho > T_1$), by Lemma \ref{Integrability}, the candidate value function
in (\ref{gvfunntc}) satisfies Condition (iii) in Theorem 
5.1 of \cite{oksendal2019applied}.

It remains to prove that the candidate value function in (\ref{gvfunntc}) satisfies Condition (iv) and Condition (vi) in Theorem 5.1 of \cite{oksendal2019applied}. We shall first prove Condition (vi). To this end, as in \cite{oksendal2019applied}, (Page 64 therein), it suffices to prove that for all $\tau \le \tau_{\cal O}$, 
\begin{eqnarray} \label{UICondition}
\mathbb{E}^{(s, x, p)} [ e^{-2 \rho \tau} (P (\tau))^{2A} (X (\tau))^{2B}] \le K^{\dagger}_1,
\end{eqnarray}
for some finite constant $K^{\dagger}_1 > 0$.

Similarly to the proof of Lemma \ref{Integrability}, by applying It\^o's differentiation rule,
it can be shown that
\begin{eqnarray} \label{P2AX2B1}
(P (t))^{2A} (X (t))^{2B} &=& (P (0))^{2A} (X (0))^{2B}
+ \int^{t}_{0} (P (u))^{2A} (X (u))^{2B} T^{\dagger}_1 d u \nonumber \\
&& + \int^{t}_{0} (P (u))^{2A} (X (u))^{2B} T^{\dagger}_2 d B_1 (u) \nonumber \\
&& + \int^{t}_{0} (P (u))^{2A} (X (u))^{2B} T^{\dagger}_3 d B_2 (u) \nonumber \\
&& + 
\int^{t}_{0} (P (u-))^{2A} (X (u-))^{2B} T^{\dagger}_4
d {\tilde N}_1 (u) \nonumber \\
&& + \int^{t}_{0}  (P (u-))^{2A} (X (u-))^{2B} T^{\dagger}_5
d {\tilde N}_2 (u), \ \mbox{say},
\end{eqnarray}
where
\begin{eqnarray*}
T^{\dagger}_2 &:=& 2 B \alpha_{\pi} \sigma_S + 2 (B \alpha_k + A) \sigma_1, \nonumber\\
T^{\dagger}_3 &:=& 2 (B \alpha_k + A) \sigma_2, \nonumber\\
T^{\dagger}_4 &:=& (1 - \alpha_{\pi} \eta)^{2B} - 1, \nonumber\\
T^{\dagger}_5 &:=& (1 - \alpha_k l + \alpha_q)^{2B} - 1.
\end{eqnarray*}
For each $t \ge 0$, let $Y^{\dagger} (t) := (P (t))^{2A} (X (t))^{2B}$.
Again, similarly to the proof of Lemma \ref{Integrability}, it can be shown that for any stopping time $\tau$,
\begin{eqnarray} \label{mgfZdagger}
&& \mathbb{E}^{(s, x, p)} [ e^{- 2 \rho \tau} Y^{\dagger} (\tau)] \nonumber\\
&=& Y(0) \mathbb{E}^{(s, x, p)} \bigg [ \exp \bigg ( (- 2 \rho + T^{\dagger}_1 + \lambda_1 (\ln(1 + T^{\dagger}_4) - T^{\dagger}_4) + \lambda_2 (\ln(1 + T^{\dagger}_5) - T^{\dagger}_5) ) \tau \bigg ) \bigg ]
\nonumber\\
&& 
\times \mathbb{E}^{(s, x, p)} \bigg [ \exp \bigg ( \ln (1 + T^{\dagger}_4) {\tilde N}_1 (\tau) \bigg ) \bigg ] \mathbb{E}^{(s, x, p)} \bigg [ \exp \bigg ( \ln (1 + T^{\dagger}_5) {\tilde N}_2 (\tau) \bigg ) \bigg ]. \nonumber\\
\end{eqnarray}
For each $n = 1, 2, \ldots$, let $\tau_{\cal O} (n) := \min (n, \tau_{\cal O})$. Then $\tau_{\cal O} (n) < \infty$, $\mathbb{P}$-a.s., for each $n = 1, 2, \ldots$, As in \cite{oksendal2019applied},
Exercise 1.6 therein, it can be shown that for all $\tau \le \tau_{\cal O} (n)$,
\begin{eqnarray} \label{mgfN1dagger}
\mathbb{E}^{(s, x, p)} \bigg [ \exp \bigg ( \ln (1 + T^{\dagger}_4) {\tilde N}_1 (\tau) \bigg ) \bigg ] = \mathbb{E}^{(s, x, p)} \bigg [ \exp \bigg ( (T^{\dagger}_4 - \ln (1 + T^{\dagger}_4) ) \lambda_1 \tau \bigg ) \bigg ] < \infty,
\end{eqnarray}
and that
\begin{eqnarray} \label{mgfN2dagger}
\mathbb{E}^{(s, x, p)} \bigg [ \exp \bigg ( \ln (1 + T^{\dagger}_5) {\tilde N}_2 (\tau) \bigg ) \bigg ] = \mathbb{E}^{(s, x, p)} \bigg [ \exp \bigg ( (T^{\dagger}_5 - \ln (1 + T^{\dagger}_5) ) \lambda_2 \tau \bigg ) \bigg ] < \infty. 
\end{eqnarray}
Consequently, using Eq. (\ref{mgfZdagger}), Eq. (\ref{mgfN1dagger}) and Eq. (\ref{mgfN2dagger}), we have:
\begin{eqnarray} \label{mgfZdagger1}
\mathbb{E}^{(s, x, p)} [ e^{- 2 \rho \tau} Y^{\dagger} (\tau)] = Y(0) \mathbb{E}^{(s, x, p)} \bigg [ \exp \bigg ( (- 2 \rho + T^{\dagger}_1 ) \tau \bigg ) \bigg ], \quad \mbox{for all $\tau \le \tau_{\cal O} (n)$}.
\end{eqnarray}
By sending $n \to \infty$, Eq. (\ref{mgfZdagger1}) becomes:
\begin{eqnarray} \label{mgfZdagger2}
\mathbb{E}^{(s, x, p)} [ e^{- 2 \rho \tau} Y^{\dagger} (\tau)] = Y(0) \mathbb{E}^{(s, x, p)} \bigg [ \exp \bigg ( (- 2 \rho + T^{\dagger}_1 ) \tau \bigg ) \bigg ], \quad \mbox{for all $\tau \le \tau_{\cal O}$}.
\end{eqnarray}
Therefore, under Condition (i) of this theorem, (say $\rho > \frac{T^{\dagger}_1}{2}$),  
\begin{eqnarray} \label{UI}
\mathbb{E}^{(s, x, p)} [ e^{- 2 \rho \tau} Y^{\dagger} (\tau)] \le K^{\dagger}_1,
\end{eqnarray}
for some finite positive constant $K^{\dagger}_1$. This proves Eq. (\ref{UICondition}). Consequently, Condition (vi) in Theorem 5.1 of \cite{oksendal2019applied} is satisfied.

We now prove that Condition (iv) in Theorem 5.1 of \cite{oksendal2019applied} is satisfied. Again, as in 
\cite{oksendal2019applied}, Page 64 therein, it suffices to prove that for all $\tau \le \tau_{\cal O}$,
\begin{eqnarray} \label{EV-}
\mathbb{E}^{(s, x, p)} [ (V^- (R(\tau)) )^2 ] \le K^{\dagger}_2, 
\end{eqnarray}
for some finite positive constant $K^{\dagger}_2$, where $V^-$ is the negative part of $V$.

\noindent 
Note that for any $\tau \le \tau_{\cal O}$,
\begin{eqnarray} \label{V-}
(V^- (R(\tau)))^2 &\le& (V^- (R(\tau)) + V^+ (R(\tau)))^2 = | V (R(\tau)) |^2 \nonumber\\
&=& \bigg ( \frac{1} {1 - \gamma} \ap_v \bigg )^2 e^{-2 \rho s} e^{- 2\rho \tau} (P (\tau))^{2A} (X (\tau))^{2B}, 
\end{eqnarray}
where $V^+$ is the positive part of $V$. 

Consequently,
\begin{eqnarray} \label{EV-1}
\mathbb{E}^{(s, x, p)} [ (V^- (R(\tau)))^2 ] \le \bigg ( \frac{1} {1 - \gamma} \ap_v \bigg )^2 e^{- 2 \rho s} \mathbb{E}^{(s, x, p)} [e^{- 2 \rho \tau} (P (\tau))^{2A} (X (\tau))^{2B}]. 
\end{eqnarray}
Since $( \frac{1} {1 - \gamma} \ap_v )^2 e^{- 2 \rho s}$ is finite, to prove (\ref{EV-}), it remains to prove that for all $\tau \le \tau_{\cal O}$,
\begin{eqnarray} \label{EV-2}
\mathbb{E}^{(s, x, p)} [e^{- 2 \rho \tau} (P (\tau))^{2A} (X (\tau))^{2B}] \le K^{\dagger}_3,
\end{eqnarray}
for some finite positive constant $K^{\dagger}_3$. 

\noindent
The inequality in (\ref{EV-2}) follows directly from (\ref{UI}). This proves that Condition (iv) in Theorem 5.1 of \cite{oksendal2019applied} is satisfied. This completes the proof.
\end{proof}

\section{Proof of the HJB function concavity}\label{ConcavityProof_sec}
\noindent In this section we prove that the following function:
\begin{align} \label{Hfunction}
H(\ap_c, \alpha_{\pi},\ap_k,\ap_{q})=& 
\frac{1}{1-\gamma}\big(\ap_{c}^{\beta}\ap_k^{1-\beta}\big)^{1-\gamma} 
+\frac{\ap_v}{1-\gamma}\Big(\beta(1-\gamma)\mu_P-\rho-\frac{1}{2}\sigma_{P}^{2}\beta(1-\gamma)(1-\beta(1-\gamma))\Big)\nonumber \\
& +\ap_v\Big((r+(1-\beta(1-\gamma))\sigma_{P}^{2}-\mu_{P})(1-\ap_k)-\delta\ap_k-\varphi\lmb_2\ap_q \nonumber\\
&
+\alpha_{\pi}\left(\mu_{S}+\lmb_1\eta-r-(1-\beta(1-\gamma))\sigma_{S}\sigma_{1}\right)-\ap_{c}\Big) \nonumber\\
& - \gamma\ap_v\Big(\frac{1}{2}\alpha_{\pi}^{2}\sigma_{S}^{2}+\frac{1}{2}\sigma_{P}^{2}(1-\ap_k)^{2}-\alpha_{\pi}\sigma_{S}\sigma_{1}(1-\ap_k)\Big) \nonumber\\
&
+\frac{\ap_v}{1-\gamma}
\Big(\lmb_1((1-\eta\alpha_{\pi})^{1-\gamma}-1)
+\lmb_2((1-\ell\ap_k+\ap_q)^{1-\gamma}-1)\Big)
\end{align}
in HJB equation (\ref{hjba}) is concave.
Note that for our model setup we have $\gamma> 0$, $\gamma\ne 1$, $0<\beta<1$.

To prove that the first term on the right-hand side of (\ref{Hfunction})  is concave, we prove that $g(\alpha_c,\alpha_k)=-\frac{(\alpha_c^\beta \alpha_k^{1-\beta})^{1 - \gamma}}{1-\gamma}$ is convex. To this end, we proceed by calculating the respective Hessian matrix with second derivatives and demonstrating that it is positive definite.
\begin{align*}
 &\frac{\partial^2 g(\alpha_c,\alpha_k)}{\partial \alpha_c^2}=\alpha_k^{(1-\beta)(1-\gamma)}\alpha_c^{\beta(1-\gamma)-2}\beta(1+\beta\gamma-\beta)>0,\\
 &\frac{\partial^2 g(\alpha_c,\alpha_k)}{\partial \alpha_k^2}=\alpha_k^{(1-\beta)(1-\gamma)-2}\alpha_c^{\beta(1-\gamma)}(1-\beta)(\beta+\gamma(1-\beta))>0,\\
 &\frac{\partial^2 g(\alpha_c,\alpha_k)}{\partial \alpha_c\partial \alpha_k}=\frac{\partial^2 g(\alpha_c,\alpha_k)}{\partial\alpha_k\partial\alpha_c}
 =\alpha_k^{(1-\beta)(1-\gamma)-1}\alpha_c^{\beta(1-\gamma)-1}\beta(1-\beta)(\gamma-1).
\end{align*}
The Hessian matrix is positive definite because all its upper left determinants are positive: in the above it is shown that $\frac{\partial^2 g(\alpha_c,\alpha_k)}{\partial \alpha_c^2}>0$, and it is easy to calculate the determinant of the full Hessian matrix
\begin{equation}
\frac{\partial^2 g(\alpha_c,\alpha_k)}{\partial \alpha_c^2}\frac{\partial^2 g(\alpha_c,\alpha_k)}{\partial \alpha_k^2}-\left(\frac{\partial^2 g(\alpha_c,\alpha_k)}{\partial \alpha_c\partial \alpha_k}\right)^2=\frac{\alpha_c^{2\beta(1-\gamma)}\alpha_k^{2(1-\beta)(1-\gamma)}}{\alpha_c^2\alpha_k^2}\beta(1-\beta)\gamma>0.
\end{equation}
Then using the facts that the concave function of linear function of arguments is concave, and the sum of concave functions is concave (well known operations preserving concavity; see e.g. \citet[section 3.2]{boyd2004convex}), it is trivial to show that function $H(.)$ is concave. 

~

\noindent The \textbf{first order derivatives} of $H(.)$ are

\begin{align}
\hspace{-1cm} \frac{\partial H}{\partial \alpha_c} = &\beta\alpha_k^{(1-\beta)(1-\gamma)}\alpha_c^{\beta(1-\gamma)-1} -\alpha_v\label{dHdalphac_eq}\\
\frac{\partial H}{\partial \alpha_{\pi}}=&
\alpha_v\big(\mu_{S}+\lmb_1\eta-r -\gamma\alpha_{\pi}\sigma_{S}^{2}-(1-\beta)(1-\gamma)\sigma_{S}\sigma_{1}-\eta\lmb_1(1-\eta\alpha_{\pi})^{-\gamma}-\alpha_k\gamma \sigma_{S}\sigma_{1}\big)\label{dHdalphapi1_eq}\\
\frac{\partial H}{\partial \alpha_k}=&
(1-\beta)\alpha_k^{(1-\beta)(1-\gamma)-1}\alpha_c^{\beta(1-\gamma)}\nonumber\\
&-\alpha_v\big(\sigma_{P}^{2}((1-\beta)(1-\gamma)+\gamma\ap_k)+\gamma\alpha_{\pi}\sigma_{S}\sigma_{1}+r-\mu_{P}+\delta +\ell\lmb_2(1-\ell\ap_{k}+\ap_{q})^{-\gamma}\big)\label{dHdalphak_eq}\\
\frac{\partial H}{\partial \alpha_q}=&\alpha_v\lambda_2 \left((1-\ell \alpha_k + \alpha_q)^{-\gamma}-\varphi \right)\label{Halphaq_eq}
\end{align}
These first-order derivatives of $H$ will be used to discuss the Karush-Kuhn-Tucker conditions and their uses to derive the solvency conditions in \eqref{param_constraints} and the system of equations in \eqref{optc} in the sequel.

~

\noindent\textbf{Karush-Kuhn-Tucker conditions}\\
\noindent For convex function and convex inequality constraints, \emph{Karush-Kuhn-Tucker conditions} are sufficient conditions to be satisfied at supremum of $H$. Taking parameterisation and formulation of KKT condition for minimisation problem as in \citet[Section 5.5.3, p.244]{boyd2004convex}, considering  constraint $\alpha_q\ge 0$, and assuming solvency positivity constraints are satisfied we have

\begin{equation}
{\bf{x}}:=(\ap_c, \alpha_{\pi},\ap_k,\ap_{q})^\top,\;f_0({\bf x}):=-H(\ap_c, \alpha_{\pi},\ap_k,\ap_{q}),\; f_1({\bf x}):=-\alpha_q,
\end{equation}
and the KKT conditions are
\begin{align}
f_1(x):=-\alpha_q\le 0,\\
\tilde\lambda_1\ge 0, \label{lambda1constraint}\\
\tilde\lambda_1 f_1(x)=0,\label{product_consition}\\
\frac{\partial f_0}{\partial \alpha_c}=0,\label{firstorder_alphac}\\
\frac{\partial f_0}{\partial \alpha_{\pi}}=0,\label{firstorder_alphapi}\\
\frac{\partial f_0}{\partial \alpha_k}=0,\label{firstorder_alphak}\\
\frac{\partial f_0}{\partial\alpha_q} +\tilde\lambda_1\frac{\partial f_1}{\partial\alpha_q}=0.\label{firstorder_alphaq}
\end{align}
Substituting (\ref{Halphaq_eq}) into the last equation of (\ref{firstorder_alphaq}), we obtain the following equality:
$$
    -\alpha_v\lambda_2 \left((1-\ell \alpha_k + \alpha_q)^{-\gamma}-\varphi \right)-\tilde\lambda_1=0.
$$
This, together with the inequality in \eqref{lambda1constraint}, lead to:
\begin{eqnarray} \label{lambda1constraint1}
\tilde\lambda_1=\lambda_2 \alpha_v(\varphi-(1-\ell \alpha_k + \alpha_q)^{-\gamma})\ge 0.
\end{eqnarray}
Thus, condition (\ref{product_consition}) becomes
\begin{equation} \label{product_consition1}
(\varphi-(1-\ell \alpha_k + \alpha_q)^{-\gamma})\times \alpha_q=0,\; \mbox{where} \; \alpha_q\ge 0.
\end{equation}
From \eqref{lambda1constraint1} and \eqref{product_consition1}, we conclude that
\begin{itemize}
\item[i)] if $\alpha_q=0$, then $\varphi-(1-\ell \alpha_k + \alpha_q)^{-\gamma}=\varphi-(1-\ell \alpha_k)^{-\gamma}\ge 0$, i.e. $\varphi^{-1/\gamma}-1+\ell \alpha_k\le 0$. Note  that for this case, 
 solvency condition $1-\ell\alpha_k+\alpha_q>0$ in \eqref{param_constraints} is satisfied because $1-\ell \alpha_k\ge \varphi^{-1/\gamma}>0$, recalling that $\varphi\ge 1$.
\item[ii)] if $\alpha_q>0$, then $\varphi-(1-\ell \alpha_k + \alpha_q)^{-\gamma}=0$ leading to $\alpha_q=\varphi^{-1/\gamma}-1+\ell \alpha_k>0$. Note that for this case, solvency condition $1-\ell\alpha_k+\alpha_q>0$ in \eqref{param_constraints} is satisfied as well because $1-\ell \alpha_k+\alpha_q=\varphi^{-1/\gamma}>0$.
\end{itemize}
The above cases (i) and (ii) can be written in a compact form as follows:
\begin{equation}\label{alphaq_KKT_eq}
\alpha_q=\max(\varphi^{-1/\gamma}-1+\ell \alpha_k,0).
\end{equation}

This equation (\ref{alphaq_KKT_eq}), together with first-order conditions (\ref{firstorder_alphac},\ref{firstorder_alphapi},\ref{firstorder_alphak}), lead to the system of equations (\ref{optc}).

We also note from (\ref{dHdalphac_eq}), 
(\ref{dHdalphak_eq}) and (\ref{dHdalphapi1_eq}) that $\frac{\partial H}{\partial \alpha_c}\rightarrow +\infty$, $\frac{\partial H}{\partial \alpha_k}\rightarrow +\infty$ and $\frac{\partial H}{\partial \alpha_{\pi}}\rightarrow -\infty$ when $\alpha_c\rightarrow 0+$, $\alpha_k\rightarrow 0+$ and $1-\eta\alpha_{\pi}\rightarrow 0+$ respectively. Thus, concave function $H$ will never have supremum at $\alpha_c=0$ or $\alpha_k=0$ or $\alpha_{\pi}=1/\eta$.

\section{Transversality conditions}\label{Transversality conditions}
Our verification theorem condition $\rho>T_1$ (from Lemma \ref{Integrability} required in Theorem \ref{verification}), i.e. in the case of $\gamma<1$, simplifies to the transversality condition in \cite{damgaard2003} in the limit of no jumps and no insurance and transversality condition in \cite{merton1969lifetime} in the limit of no jumps and no durable goods.
In the case $\gamma>1$ of Theorem \ref{verificationCand2} we additionally require $\rho > \frac{T_1^\dagger}{2}$. 

In the case of no jumps in durable goods and no insurance, and no jumps of risky asset we have $\lambda_2=0$, $\lambda_1=0$, $\eta=0$, $\ell=0$, $\alpha_q=0$ and thus  condition $\rho > \frac{T_1^\dagger}{2}$ reduces to

\beqr{}
\rho>\frac{1}{2}T^\dagger_1 =&  B (r + \ap_{\pi} (\mu_S  - r) + \ap_k
(\mu_P - r - \delta) - \ap_c) + 
\frac{1}{2}B(2B - 1) ((\ap_{\pi} \sigma_S + \ap_k \sigma_1)^2 + \ap^2_k \sigma^2_2) \\
&+ A \bigg (\mu_P + \frac{1}{2} (2A - 1) (\sigma^2_1 + \sigma^2_2) \bigg )
+ 2AB(\ap_{\pi} \sigma_S \sigma_1 + \ap_k (\sigma^2_1 + \sigma^2_2)),
\eeqr 
and condition  $\rho > T_1$ reduces to
\beqr{tildeT1}
\rho>{T}_1 =& B (r + \ap_{\pi} (\mu_S - r) + \ap_k
(\mu_P - r - \delta) - \ap_c) + 
\frac{1}{2}B(B - 1) ((\ap_{\pi} \sigma_S + \ap_k \sigma_1)^2  + \ap^2_k \sigma^2_2) \\
&+ A \bigg (\mu_P + \frac{1}{2} (A - 1) (\sigma^2_1 + \sigma^2_2) \bigg )
+ AB(\ap_{\pi} \sigma_S \sigma_1 + \ap_k (\sigma^2_1 + \sigma^2_2)). 
\eeqr 

These two conditions are different and correspond to cases $\gamma>1$ and $\gamma<1$ respectively, while in \cite{damgaard2003} there is only one condition for the case of $\gamma<1$.

In fact, it is easy to check that second condition $\rho>{T}_1$ is indeed the transversality condition derived in \cite{damgaard2003}; see Appendix 3, page 237 in \cite{damgaard2003} where $\xi$ corresponds to our $T_1$. Here we note that only $\gamma\in(0,1)$ case is considered in \cite{damgaard2003} and \cite{matos2014}.

~

Let us check whether our conditions reduce to the transversality condition in the Merton problem. 
The Merton problem transversality condition (i.e., no jumps and no durable goods) is
$$
\rho>(1-\gamma)\left[r+\frac{(\mu_S-r)^2}{2\sigma_S^2\gamma}\right]
$$
}
with optimal $\alpha_{\pi}=\frac{\mu_S-r}{\gamma\sigma_S^2}$ 
and 
$$\alpha_c=\frac{1}{\gamma}\left(\rho-(1-\gamma)r-\frac{(1-\gamma)(\mu_S-r)^2}{2\sigma_S^2\gamma} \right).$$

In this limit we have $\beta=1,\alpha_k=0, A=0,\eta=0,\ell=0,\lambda_1=0,\lambda_2=0,\alpha_q=0$ and our conditions, after substituting the above optimal $\alpha_{\pi}$ and $\alpha_c$, reduce to
\begin{align}
\rho>\frac{1}{2}T^\dagger_1 =&  B (r + \ap_{\pi} (\mu_S  - r)  - \ap_c) + 
\frac{1}{2}B(2B - 1) (\ap_{\pi} \sigma_S)^2\nonumber \\
=&(1-\gamma)\left(r+\frac{(\mu_S-r)^2}{2\gamma\sigma_S^2}(2-\gamma) \right)
\end{align}
and
\begin{align}
\rho>{T}_1 =& B (r + \ap_{\pi} (\mu_S - r)  - \ap_c) + 
\frac{1}{2}B(B - 1) (\ap_{\pi} \sigma_S )^2\nonumber \\
=&(1-\gamma)\left(r+\frac{(\mu_S-r)^2}{2\gamma\sigma_S^2} \right).
\end{align} 
The last one is transversality condition in the case of Merton problem. We also see in this limiting case that 
$\frac{1}{2}T^\dagger_1={T}_1+\frac{1}{2} B^2\alpha_\pi^2 \sigma_S^2>{T}_1$, i.e., the condition $\rho>T^\dagger_1$ in the case $\gamma>1$ is more conservative than the transversality condition in the case $\gamma<1$.
\end{document}

%% file: newcommands.tex
\usepackage{graphicx}
\usepackage{amssymb}
\usepackage{bbm}
\usepackage{amsmath}

\newcommand{\qh}{\hat{q}}
\newcommand{\ap}{\alpha}

\newtheorem{lem}{Lemma}

\algnewcommand\INPUT{\item[\textbf{Input:}]}%
\algnewcommand\OUTPUT{\item[\textbf{Output:}]}%

\makeatletter

\numberwithin{equation}{section}

  \theoremstyle{remark}
  
  \theoremstyle{plain}
  \newtheorem{thm}{\protect\theoremname}[section]
  \theoremstyle{plain}

\makeatother

\providecommand{\propositionname}{Proposition}
\providecommand{\remarkname}{Remark}
\providecommand{\theoremname}{Theorem}

\newcommand{\Prt}[1]{\left({#1}\right)}

\newcommand{\pd}{\partial}
\newcommand{\beql}[1]{\begin{equation}\label{#1}}
\newcommand{\eeq}{\end{equation}}
\newcommand{\beqr}[1]
{\begin{equation}\label{#1}\begin{aligned}}
\newcommand{\eeqr}
{\end{aligned}\end{equation}}
\newcommand{\beqc}[1]
{\begin{equation}\label{#1}\begin{cases}}
\newcommand{\eeqc}
{\end{cases}\end{equation}}
\newcommand{\bry}[1]
{\begin{array}{#1}}
\newcommand{\ery}{\end{array}}

\newcommand{\lmb}{\lambda}
